\documentclass[pra,twocolumn,notitlepage,superscriptaddress,longbibliography]{revtex4-1}
\usepackage{amsmath}
\usepackage{amssymb}
\usepackage{mathrsfs}
\usepackage[english]{babel}

\usepackage[unicode=true,colorlinks=true,citecolor=blue,urlcolor=blue]{hyperref}

\usepackage{bm}
\usepackage{color}
\usepackage{epsfig}

\usepackage[normalem]{ulem}

\renewcommand {\phi}{{\varphi}}

\begin{document}

\title{Quantum Optical Signatures of Band Geometry in Solid-State High Harmonics}

\author{Denis Ilin}
\affiliation{School of Mathematical and Physical Sciences, University of Technology Sydney, Ultimo, NSW 2007, Australia}\affiliation{ Sydney Quantum Academy, Sydney, NSW 2000, Australia}

\author{Alexander S. Solntsev}
\affiliation{School of Mathematical and Physical Sciences, University of Technology Sydney, Ultimo, NSW 2007, Australia}

\author{Ivan Iorsh}
\affiliation{Queen's University, Kingston, Canada}

\begin{abstract}
We develop a general theory of high-harmonic generation (HHG) in solid-state systems, based on a weak-correlation expansion of photonic and matter degrees of freedom. Unlike standard HHG theories, which treat light-matter dynamics through the Schrödinger equation, our approach employs density-matrix evolution, naturally capturing the mixed-state character of both the field and the matter - a critical aspect for describing complex solid-state band structures. We show explicitly that the properties of the emitted fields are governed by the quantum statistics and quantum geometry of the underlying solid. Taking the Su-Schrieffer-Heeger (SSH) model in a one-sided optical cavity as a paradigmatic example, we demonstrate the change in HHG response and quantum-light signatures due to the presence of the cavity. Furthermore, we show that cavity-matter interaction gives rise to squeezed high-harmonic quantum light, whose properties are directly imprinted by the current-current fluctuations and conductance in the material system. Crucially, the observed squeezing does not rely on a separate quartic Kerr mechanism. In the mesoscopic regime, the genuine quantum Kerr term is higher order in the light-matter coupling strength and negligible, while the relevant non-classical effect is governed by current-current fluctuations encoded in the complex susceptibility of the material. Additionally, non-trivial cross-correlations between different far-field emitted modes appear and show bunched and anti-bunched mode pairs. This work establishes a direct link between band geometry and photon statistics, opening new avenues for geometry-sensitive quantum light generation and photon statistics based spectroscopy of solid-state systems.

\end{abstract}

\maketitle

\section{Introduction}
High-harmonic generation (HHG) is an extreme nonlinear optical phenomenon that arises when matter—whether isolated atoms, molecules, or condensed solids—is driven by an intense laser and emits radiation at integer multiples of the driving frequency~\cite{PhysRevLett.71.1994, 10.1063/1.89628, Ferray_1988}. Over decades of experiment and theory, HHG has become the main framework for coherent extreme-ultraviolet sources and the foundation of attosecond science~\cite{doi:10.1126/science.1059413, RevModPhys.81.163}. In dilute gases the process is commonly rationalized by the intuitive three-step picture~\cite{PhysRevA.49.2117, PhysRevLett.68.3535}: under a strong field an electron escapes the suppressed atomic potential, it is accelerated by the laser in the continuum, and—when conditions permit—returns to recombine with the parent ion while releasing a high-energy photon. Repetition each optical cycle produces the characteristic comb-like harmonic spectrum. In solids~\cite{PhysRevLett.113.073901, PhysRevLett.107.167407, Schubert2014,Ghimire}, HHG is often described by a combination of interband and intraband currents: interband emission admits an analog of the three-step mechanism (promotion across a bandgap, acceleration of electron and hole, and recombination), while intraband emission stems from field-driven motion within non-parabolic bands; both contributions are in general coupled and shape the observed spectrum.

Traditional theoretical treatments typically quantize the matter but treat both the driving laser and the radiated field classically, modeling emission as dipole radiation given by the expectation value of the driven dipole. Semiclassical approaches—exemplified by Lewenstein-type semi-analytic theories \cite{PhysRevA.49.2117, PhysRevLett.113.073901} and refined by many ab initio calculations—have achieved striking agreement with HHG spectra from atoms and molecules. Nevertheless, these hybrid descriptions are limited in scope: they do not capture inherently quantum properties of the electromagnetic field nor all back-action between light and matter. 

When the field is treated quantum mechanically, HHG reveals new layers of behavior~\cite{Gonoskov, PhysRevA.101.013418, PhysRevA.101.013410, Gorlach2020, lange2025highorder}. Quantum optical theory and experiments have shown that harmonic emission can exhibit non-classical photon statistics, mode correlations, and entanglement between light and the generating medium. Transitions connecting different initial and final electronic states can produce multiple spectrally shifted harmonic combs; the emitted modes may depart from Poissonian statistics and display squeezing ~\cite{PhysRevLett.132.143603}, bunching, or correlated two-mode behavior~\cite{PhysRevA.110.063118}; and corrections beyond the dipole approximation can give rise to even harmonics even for a monochromatic driver~\cite{Ciappina}. Importantly, conditional protocols—measurements or photon subtraction on selected modes—can herald non-classical states such as coherent-state superpositions (cat states)~\cite{PhysRevX.15.011023}, offering routes to quantum-state engineering within strong-field platforms~\cite{PhysRevLett.130.166903, Pizzi, Kim}.

These quantum features become richer in systems with strong electron–electron interactions. Correlated materials including Mott insulators and other beyond-mean-field phases modify electronic dynamics and can enhance non-classical light generation~\cite{PhysRevLett.129.157401, PhysRevResearch.3.023250, PhysRevLett.128.127401}. Modeling such effects with canonical many-body Hamiltonians (for example, the Fermi–Hubbard model) reveals that increasing correlations can qualitatively alter the photon statistics and squeezing of the emitted field~\cite{PhysRevA.109.033110, PhysRevA.111.013113}. Furthermore, preparing ensembles in correlated or collective states maps many-body correlations onto photonic observables, enabling entanglement and squeezed-light emission not present in uncorrelated phases.

Despite this progress, existing quantum-optical HHG theories have mainly focused on matter systems treated effectively at the pure-state level, i.e. composite light-matter systems that can be described via evolution of a wavefunction. For thermally occupied multiband solids~\cite{PhysRevB.103.125419, zbfh-qfd2, Kim} and reduced light-matter dynamics, however, a density-matrix formulation is more natural, as it accommodates mixed-state occupations, dephasing, and band-resolved correlations without restricting the description to a single wavefunction. At the same time, several theories of second harmonic generation within the quantum kinetic framework have been established~\cite{PhysRevB.61.5337, PhysRevLett.129.227401, PhysRevB.97.085201}, where particular attention has been paid to different quantum geometric properties of the electronic wave-function in semiconductors. To our knowledge, a general framework that combines a quantum-kinetic description of a solid with predictions for the quantum statistics of the emitted HHG field has not yet been established.

In this paper, we develop a formalism that connects the material properties, such as geometry and electric current and dynamical current-current fluctuations induced by a multi-mode coherent-state laser, with the field-counting photonic statistics of the emitted field in the problem of HHG. This formalism is presented via a universal Lindblad-type (ULL) equation~\cite{PhysRevX.10.041024} written for the reduced density matrices of the quantum field degrees of freedom and a solid-state system that takes into account the weak correlation mechanism between the  emitted field and the matter system. The main idea is based on a rotation of the initial photon space onto the external laser field, which in turn decouples the system into two subsystems, namely a classically pumped material structure and a photon vacuum state \cite{PhysRevA.111.013113}. Due to the weak light-matter interaction, general light-matter correlations can be studied perturbatively, where the dominant semiclassical contribution arises from a classically driven matter system that produces coherent fields, and there is a weak back-action force of the emitted field on the material system due to the mesoscopic interaction regime. The nature of each contribution in the total emitted field is studied in detail. We demonstrate that the current–current fluctuations arising from the coupling between the solid-state emitter and the photonic vacuum dictate the quantum properties of the emitted light, giving rise to multi-mode correlations and squeezing. Notably, in the mesoscopic regime the enhanced by the classical field Kerr nonlinearity is incorporated in the quantum fluctuations mediated by the susceptibility of the system. Crucially, the careful description of the matter correlation functions reveals boosted resonant emission and enhanced quantum features in the presence of the cavity. We emphasize the potential benefits of using the cavity setups to generate enhanced quantum light in HHG. As a concrete example, the one-dimensional Su–Schrieffer–Heeger (1D SSH) model is considered. The universality of our formalism allows it to be applied to any type of matter system, namely solids, atoms, or gases. Our study shows how to analyze the material geometric properties from studying the photonic statistics, or, vice versa, how to generate quantum light from the geometry of a mesoscopic system. In this paper atomic units $(\hbar= m_e=-e = 4\pi\epsilon_0 = 1)$ are used unless explicitly stated otherwise.

The rest of this paper is organized as follows. In Sec.~\ref{sec:summary} we provide a brief summary of the main physical results. In Sec.~\ref{sec:general_model} we define the general model and derive the reduced density matrix equation for the photonic system. In Sec.~\ref{sec:field_statistics} we obtain equations of motion for the photonic field operators and then we link far field in Sec.~\ref{sec:far_field} cavity statistics in the one-side cavity setup and current-current correlation in the solid-state system. In Sec.~\ref{sec:1d_ssh} we consider 1D SSH model inside the cavity as an example of the developed theory. We conclude with an outlook in Sec.~\ref{sec:conclusion}.

\section{Summary of results}
\label{sec:summary}

We start with a general electronic crystal structure driven by a multi-mode coherent-state laser, coupled weakly to a quantized electromagnetic field. We decompose the vector potential into a classical part which drives the material system semiclassically and a quantum part which is responsible for the emitted radiation. This rotation maps the initial laser field to a vacuum state, so that only the field emitted by the crystal is tracked. The total Hamiltonian is then expanded to second order in the light-matter coupling strength $g_0$, yielding an effective Hamiltonian Eq.~\eqref{Eq:general} that contains both paramagnetic and diamagnetic coupling terms, the latter encoding a Kerr-like nonlinearity enhanced by the classical driving field. 

Because the total matter-field state cannot in general be described as a pure state, we formulate a universal Lindblad-like (ULL) master equation for the coupled reduced density matrices of the photonic and material subsystems Eq.~\eqref{Eq:weak_coup_eq}. The weak-coupling regime $N^2g_0\ll1$ permits a controlled expansion. To leading order in this expansion, the material density matrix is shown to remain close to its initial thermal equilibrium state Eq.~\eqref{Eq:rhoM}, with corrections driven by the average quantum field. This allows us to express the total paramagnetic current as a sum of the classically induced current and a back-action correction mediated by the nonlinear dynamical susceptibility $\chi_{\mu,\nu}(t,t')$ in Eq.~\eqref{Eq:current}. The back-action term reflects the retarded influence of previously emitted fields on the material, and represents the leading quantum correction to the semiclassical HHG current.

The reduced photonic density matrix obeys a Lindblad equation Eq.~\eqref{Eq:generalULL} in which the Lindblad dissipator is governed by the two-time current-current correlation tensor $D_{\mu,\nu}(t,t')$ of the driven material system. This tensor, evaluated over the initial equilibrium state, encodes key quantum geometric properties of the solid, its antisymmetric part coincides with the nonlinear susceptibility of the matter system. We emphasize that its non-vanishing value Eq.~\eqref{Eq:zeroD} reflects intrinsic quantum current fluctuations in the crystal, which are the origin of non-classical light generation.

Turning to specific observables, we write the equation of motion for the intracavity field amplitude Eq.~\eqref{Eq:amplitude} and identify two contributions: a classical semiclassical term grows as $g_0 N$, and a quantum back-action correction grows as $g_0^3 N^{2}$ in Eq.~\eqref{Eq:exp_amp}, which is always beyond of the consideration due to the natural weak light-matter coupling. The far-field power spectrum is decomposed into a coherent part and a quantum noise contribution in Eq.~\eqref{Eq:anal_intens}, where the latter is strictly positive and is governed by the spectral noise function $S_{\mu,m}(t)$, which is proportional to the Fourier transform of $D_{\mu,\nu}$ evaluated at the cavity resonance frequency Eq.~\eqref{Eq:quan_pop}. The quantum noise contribution scales as $g_0^2N$ and vanishes in the classical $N \to \infty$, as expected. Moreover, we find that the cavity works as a filter for far-field statistics, which means it boosts the resonant emission by suppressing the non-resonant scattering.

For single-mode far-field photon statistics, we derive an expression for the second-order correlation function $g_2(\omega)_{\mu}$ in Eq.~\eqref{Eq:g2wans}, which is dominated by the semiclassical field and generically yields super-Poissonian statistics. We further relate the degree of squeezing of the far-field modes to the current–current fluctuations, Eq.\eqref{Eq:degree_sq}. It is governed by an anomalous correlation, which consists of two parts, namely the correlation matter tensor and the longitudinal component of the conductivity tensor in Eq.\eqref{Eq:M_anom}. For two-mode statistics, we derive a correlation function that, as in the single-mode case, is determined via a spectral filter function. All nontrivial two-mode correlations arise exclusively from the quantum noise term and are therefore sensitive to the material’s transport properties.

As a concrete example, we apply this formalism to the one-dimensional Su–Schrieffer–Heeger (1D SSH) model in its topologically trivial phase, in the presence of the photon bath. Owing to the nonzero bath temperature, the characteristic high-harmonic plateau is attenuated in Fig.~\ref{fig:1dssh}(b), because the coherence time of interband transitions becomes finite. For an SSH chain in the mesoscopic regime, the cavity setup leads to enhanced resonant scattering in the emission spectrum in Fig.~\ref{fig:spectrum}. Furthermore, every mode exhibits nonzero squeezing, which is maximal for the first few plateau modes in Fig.~\ref{fig:squeezing}. The squeezing increases with the cavity quality factor. Finally, the two-mode correlations in Fig.~\ref{fig:two-mode} reveal a rich structure: some pairs of modes are bunched, whereas others are anti-bunched.

These results demonstrate that implementing cavity provides a practical advantage for generating non-classical light via high-harmonic generation, and that measurements of photon statistics constitute a new probe of quantum geometry in solids.

\section{General model}
\label{sec:general_model}

\begin{figure}[t]
\centering
\includegraphics[width=0.48\textwidth]{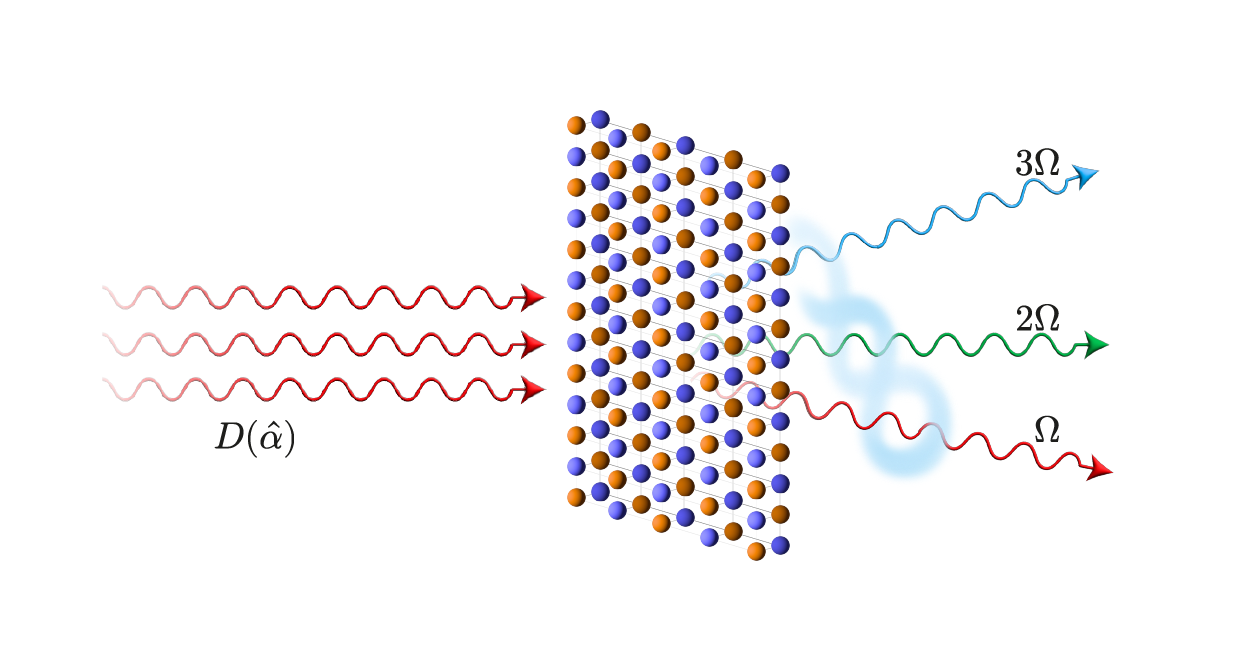}
\caption{The solid-state structure is driven by external multi-mode coherent-state laser. The relation between photonic statistics and the matter transport correlations is studied.}\label{fig:general_system}
\end{figure}

In this section, we develop a fully quantum formalism that describes the emission of quantum light as the response of an electronic system initially in thermal equilibrium under a strong time-dependent electromagnetic field (see  Fig.~\ref{fig:general_system}). The method is based on quantum electrodynamical perturbation theory that enables the study of the quantum statistics of the emitted fields. Although some initial steps involved in this theory can be found in recently developed quantum theories of the HHG \cite{Gorlach2020, PhysRevA.109.033110}, the novelty of our formalism lies in a density matrix approach that takes into account the incoherent nature of the emitted light states. The full derivation can be divided into three independent parts: first, we derive the effective time-dependent Hamiltonian; second, we formulate a coupled system of equations for the reduced material and photonic density matrices; and finally, we find the equation of motion for intracavity light observables. Although the following formalism is suitable for any photonic bath realization, in this study, we consider only a one-sided cavity setup

We start with a general electronic crystal structure that can be written as 
\begin{equation}                        
    H_{sys}=\sum\limits_{l,j}t_{l,j}\hat{c}^{\dagger}_{l}\hat{c}_{j}+..,
\end{equation}
where the indices $l,j$ encode an electron position, $t_{l,j}$ is the hopping strength between the points $l$ and $j$ and the sum accounts for more complex interaction terms. The electronic system is driven by an external laser field that is described by a multi-mode coherent state laser $|\psi_{laser}(t)\rangle=\otimes_{\textbf{q},\sigma}|\alpha_{\textbf{q},\sigma}e^{-i\omega_\textbf{q}t}\rangle$, where $\omega_q$ is the dispersion of the photonic mode with the wave vector $\textbf{q}$, $\sigma$ is the polarization and $\alpha_{\textbf{q},\sigma}$ represents the coherent states parameters. Following the Peierls substitution method, it is possible to represent the interaction of the electromagnetic field with the system as $t_{l,j}\rightarrow t_{l,j}\exp\left(i\int_{j}^{l}\hat{\textbf{A}}(\textbf{r},t)d\textbf{r}\right)$, where $\hat{\textbf{A}}$ is vector potential operator. The total Hamiltonian can then be expressed as $\hat{H}=\hat{H}_{sys-int}+\hat{H}_{F}$, where $\hat{H}_{sys-int}$ is the system Hamiltonian with the Peierls substitution and $\hat{H}_{F}$ is the Hamiltonian of electromagnetic modes. The first step of our formalism is the same as in previous quantum HHG theories, namely, we implement the rotation of the total Hamiltonian to decompose the vector potential $\textbf{A}$ into a sum of a classical time-dependent part $\textbf{A}_{cl}=\langle\psi_{laser}(t)|\textbf{A}|\psi_{laser}(t)\rangle$ and a quantum part $\textbf{A}_Q$
\begin{equation}
    \textbf{A}_Q=\sum\limits_{\textbf{q},\mu} \frac{g(\textbf{q})}{\sqrt{\omega_{\textbf{q},\mu}}}\left(\textbf{e}_{\mu}\hat{a}_{\textbf{q},\mu}e^{i\textbf{qr}}+\text{h.c.}\right),
\end{equation}
where $g(\textbf{q})$ is the light-matter coupling strength whose particular nature depends on the surrounding photonic structure, in the simple empty space case, it is $g_0=\sqrt{2\pi/V}$, where $V$ is a quantization volume, the summation $\sum_{\textbf{q}, \mu}$ is over all photonic modes with momentum $\textbf{q}$ and polarization $\mu$, $\hat{a}_{\textbf{q}, \mu} (\hat{a}^{\dagger}_{\textbf{q}, \mu})$ is an annihilation (creation) operator and  $\textbf{e}_{\sigma}$ is a unit vector of polarization. The rotation shifts the initial state of the field to the vacuum state as $\hat{D}(t)|\psi_{laser}(t)\rangle=|vac\rangle$. This effectively corresponds to a subtraction of the laser field from the total field and hence, in this picture, only the field emitted by the electronic structure is present. As a result, the laser pumping is embedded in the classical field $\textbf{A}_{cl}$ that drives the electronic structure. At the same time, the quantum field term $\textbf{A}_Q$ is responsible for the emission into space, which contains multi-mode coherent radiation in full accordance with semiclassical HHG theory and additional quantum corrections to the classically emitted state. Naturally, the light-matter interaction strength $g_0$ is small enough to serve as an expansion parameter for perturbation theory. Therefore, the total Hamiltonian can be written up to the second order in the coupling strength $g_0$ as
\begin{equation}
\begin{aligned}
    \hat{H}'(t)&=\hat{H}_{cl}(t)+\hat{H}_{ph}\\&+j^{(s)}_{p,\mu}(t)A^{(\mu)}_{Q} 
    +\frac{1}{2}j^{(s)}_{d,\mu\nu}(t)A^{(\mu)}_{Q}A^{(\nu)}_{Q},\end{aligned}\label{Eq:general}
\end{equation}
where $j^{(s)}_{p,\mu}(t)$ and $j^{d,(s)}_{d,\mu\nu}(t)$ are time-dependent paramagnetic and diamagnetic terms, respectively, the index $s$ represents the Schrödinger picture and 
the sum over the repeating polarization indices $\mu,\nu$ is implied. The first line corresponds to the electronic system under classical pumping force $\hat{H}_{cl}(t)$ and a photonic system $\hat{H}_{ph}$, while the second line represents the light-matter coupling through paramagnetic and diamagnetic responses. The explicit expressions of current operators are determined by the system Hamiltonian under the classical pumping, which manifests in time-dependent behavior. In the case where higher-order statistics of the emitted field are of interest, Eq.~\eqref{Eq:general} should contain other orders of expansion with respect to $g_0$. 

It should be noted, that Eq.~\eqref{Eq:general} is written in the velocity gauge and this choice is well motivated. Although in the length gauge it is possible to take into account all field terms simultaneously by introducing the field-matter interaction term as $\textbf{d}\textbf{E}_{Q}$, the main problem occurs in the definition of dipole matrix elements for translational-invariant systems such as crystals \cite{PhysRevLett.134.106403}. It can be shown, this singularity corresponds to the term that is proportional to the derivative of the Dirac delta function of the electronic momenta $\nabla_{\textbf{k}}\delta(\textbf{k}-\textbf{k'})$ and hence, some special techniques should be implemented to overcome this issue. Nevertheless, a relatively simple method can be used~\cite{PhysRevB.61.5337} to calculate the single-time average of observables in crystals; however, it is futile for calculating the correlation functions that arise in our study. As a result, we choose the velocity gauge and consider only the weak light-matter coupling regime.   

In this research, we use the long wavelength limit (dipole approximation), which means that we neglect spatial dispersion and assume a uniform vector potential $\textbf{A}$ over the entire crystal structure. This approximation is justified \cite{Gorlach2020}, in case we are interested in only the first few emitted harmonics. Note that, under the dipole approximation, the hopping strengths are reduced to $t_{l,j}\rightarrow t_{l,j}\exp\left(i\Delta\textbf{r}_{lj}\textbf{A}(t)\right)$. By switching to the interaction picture, we finally obtain the effective Hamiltonian of the electronic system under the classical pumping force
\begin{equation}
\begin{aligned}
    \hat{H}_{eff}(t)&=j_{p,\mu}(t)A^{(\mu)}_{Q}(t)+\frac{1}{2}j_{d,\mu\nu}(t)A^{(\mu)}_{Q}(t)A^{(\nu)}_{Q}(t),
    \end{aligned}
\end{equation}
where paramagnetic and diamagnetic terms include all orders in the classical field expansion. Hence, the Kerr-nonlinearity term enhanced by the coherent field that is quadratic in the photonic operators, i.e.$\sim(a^2+a^{{\dagger}^2})$ is incorporated in the diamagnetic current, whereas the purely quantum Kerr nonlinearity, i.e.$\sim (a^{\dagger}a)^2$ is proportional to $g^4_0$ and hence negligible.

From the semiclassical HHG theories~\cite{PhysRevA.49.2117, PhysRevLett.68.3535,PhysRevLett.113.073901} we know that the leading term in the emitted field is fully governed by paramagnetically induced currents driven by laser pumping. Besides the semiclassical current response, the system evolution determines the quantum contributions to the total emission, which destroys the coherent state nature of the emission and results in the non-classicality of photon statistics. However, the important point here is that the emitted field in the past can induce extra currents in the system due to the non-trivial conduction properties, which results in a correction to the total matter current. Such a back-action diagram arises in the second order in expansion of $g_0$ and hence must be taken into account. Initially, at $t=-\infty$ the matter subsystem and the field are decoupled, i.e. $\rho_{MF}(-\infty)=\rho_{M}(-\infty)\otimes |vac\rangle\langle vac|$, where $\rho_M(-\infty)\equiv\rho_0$ is the incoherent thermal equilibrium state of the matter subsystem. Because the light-matter interaction is relatively weak, it allows us to write down a weak-correlation expansion which takes into account the reciprocal influence of matter and emitted fields. The main idea of studying the evolution of the total matter-field state $\rho_{MF}(t)$ is its decomposition in terms of an uncorrelated part, given by the tensor product of the instantaneous reduced states of the field subsystem $\rho_F(t) = \text{Tr}_M[\rho_{MF}(t)]$ and $\rho_M(t) = \text{Tr}_F[\rho_{MF}(t)]$, and the remainder $\chi_{MF}(t)$ which carries all correlations in the total system state,
\begin{equation}
    \rho_{MF}(t)=\rho_M(t)\otimes\rho_F(t)+\chi_{MF}(t).
\end{equation}
The weak-correlation expansion for the corresponding correlation function $\chi_{MF}(t)$~\cite{PhysRevX.10.041024} requires a coupled system of equations for both matter and photonic subsystems simultaneously. This result emphasizes that the total evolution is not unitary and hence a general HHG problem of the material system cannot be solved in terms of the pure state evolution. At the beginning, the initial matter-field correlation vanishes, i.e. $\chi_{MF}(-\infty)=0$. In light of this, we use the universal Lindblad-like (ULL) dynamical master equation to determine coupled system as 
\begin{equation}
    \begin{aligned}
        \dot{\rho}_{M}(t) = &-i\left[\mathrm{Tr}_{F}[H_{eff}(t)\rho_{F}(t)\right], \rho_{M}(t)] \\
-& \mathrm{Tr}_{F}\Bigl[H_{eff}(t), \int\limits_{-\infty}^{t} dt'[\tilde{H}_{eff}(t'), \rho_{M}(t') \otimes \rho_{F}(t')]\Bigr], \\
\dot{\rho}_{F}(t) = &-i\left[\mathrm{Tr}_{M}[H_{eff}(t)\rho_{M}(t)\right], \rho_{F}(t)] \\
-& \mathrm{Tr}_{M}\Bigl[H_{eff}(t), \int\limits_{-\infty}^{t} dt'[\tilde{H}_{eff}(t'), \rho_{M}(t') \otimes \rho_{F}(t')]\Bigr],
    \end{aligned}\label{Eq:weak_coup_eq}
\end{equation}
where
\begin{equation}
\begin{aligned}
    \tilde{H}_{eff}(t')&=\hat{H}_{eff}(t')\\&-\text{Tr}_{M}[\hat{H}_{eff}(t')\rho_M(t')]-\text{Tr}_{F}[\hat{H}_{eff}(t')\rho_F(t')]
\end{aligned}\label{Eq:chi_type}
\end{equation}
We note that the obtained system of equations is written in the non-Markovian regime. The reason for this is the matter correlation function that possesses highly non-local temporal correlations.

Before proceeding further, we should address the many-body nature of the considered process. Because of the multiple emitters in the matter system, the effective light-matter interaction is increased. Hence, a new effective light-matter coupling strength must be taken into account in order to construct the perturbative expansion. In this research, we study the regime in which $N^{2}g_0\leq1$. This restriction guaranties the validity of the following results and implementation of a perturbative approach. Regarding matter observables, we note that the average paramagnetic (or diamagnetic) current, $\langle j_{p,\mu}(t)\rangle$, grows linearly with the number of emitters in the leading term, i.e. $\langle j_{p,\mu}(t)\rangle\sim N$. Hence, the semiclassical response produces the average electromagnetic field $\text{Tr}\left( A^{(\mu)}_Q(t)\rho_F(t)\right)$, the dominant contribution of which grows as $Ng^2_0$. At the same time, the current-current correlation function $\langle[j_{p,\mu}(t),j_{p,\mu}(t')]\rangle$ cannot grow faster than $N$. Altogether, these scaling laws allow us to integrate the matter density matrix equation and suppress the non-Markovian nature in the two leading terms (see Appendix~\ref{app:A1}). 
\begin{equation}
\begin{aligned}
    \rho_M(t)&=\rho_0-i\int\limits_{-\infty}^{t} dt'\left\langle A^{(\mu)}_Q(t')\right\rangle_F [j_{p,\mu}(t'),\rho_0]
    \end{aligned}\label{Eq:rhoM}
\end{equation}
where $\langle..\rangle_F=\text{Tr}(...\rho_F(t))$ and $\rho_0\equiv\rho_M(-\infty)$. From here, we can immediately calculate the paramagnetic current as 
\begin{equation}
    \langle j_{p,\mu}(t)\rangle=\langle j_{p,\mu}(t)\rangle_0+\int\limits_{-\infty}^{t}\limits dt' \chi_{\mu,\nu}(t,t')\left\langle A^{(\nu)}_Q(t')\right\rangle_F ,\label{Eq:current}
\end{equation}
where $\langle\rangle_0$ denotes averaging over the initial density matrix of the matter system, i.e. $\langle j_{p,\mu}(t)\rangle_0=\text{Tr}(j_{p,\mu}(t)\rho_0)$ is the classical current in the system induced by the laser, $\chi_{\mu,\nu}(t,t')=-i\langle[j_{p,\mu}(t),j_{p,\nu}(t')]\rangle_0$ is the nonlinear susceptibility of the solid-state system.  As can be seen, the total current in the system results from the laser pumping and from previously emitted fields, which can induce currents via nonlinear susceptibility, thereby introducing important retardation effects into the total picture. This result is in good accordance with the traditional Kubo formula, as is well known, except that the susceptibility should be calculated for the classically driven system; hence it is the dynamical susceptibility that matters for HHG problems in general. Finally, we note that the average current grows as $N(1+g^2_0)$, this aspect will be of great importance in the following field statistics study.

The main goal of this work concerns the evolution of the emitted field. Therefore, we can write the reduced density matrix equation in the Lindblad form as
\begin{equation}
\begin{aligned}
        \dot{\rho}_F(t)=&-i\left[\text{Tr}_M\left(\hat{H}_{eff}(t)\rho_M(t)\right),\rho_F(t)\right] \\&+ \int\limits_{-\infty}^{t}dt'\mathcal{L}[\rho_F(t')],   
\end{aligned}\label{Eq:generalULL}
\end{equation}
where 
\begin{equation}
    \begin{aligned}
\mathcal{L}[\rho_F&(t')]= i\chi_{\mu,\nu}(t,t')\langle A^{(\nu)}(t')\rangle_F\left[A^{(\mu)}(t),\rho_F(t')\right]\\+&D_{\mu,\nu}(t,t')\\&\times\left[A^{(\nu)}_Q(t')\rho_F(t')A^{(\mu)}_Q(t)-A^{(\mu)}_Q(t)A^{(\nu)}_Q(t')\rho_F(t')\right]\\
+&D_{\nu,\mu}(t',t)\\&\times\left[A^{(\mu)}_Q(t)\rho_F(t')A^{(\nu)}_Q(t')-\rho_F(t')A^{(\nu)}_Q(t')A^{(\mu)}_Q(t)\right],
    \end{aligned}\label{Eq:ull}
\end{equation}
where we denote $D_{\mu,\nu}(t,t')=\langle j_{p,\mu}(t)j_{p,\nu}(t') \rangle_M-\langle j_{p,\mu}(t)\rangle_M\langle j_{p,\nu}(t') \rangle_M$ is the time correlation tensor of the matter degrees of freedom and $\langle O(t)\rangle_M=\text{Tr}_{M}(O(t)\rho_M(t))$ is the averaging over the matter subsystem. A careful consideration of the asymptotics reveals that it is possible to average over the material subsystem in the Lindblad term over $\rho_0$ to obtain the dominant contribution. Hence, in the following, the correlation tensor is calculated via $\rho_M(t)\rightarrow\rho_0$  (see Appendix~\ref{app:A2}). By solving it, all the necessary statistics of the quantum field can be obtained. We note that the correlation matter tensor $D_{\mu,\nu}(t,t')$ is a more general object than susceptibility, because $\chi_{\mu,\nu}=-i(D_{\mu,\nu}-D_{\nu,\mu})$. Hence, it is natural to anticipate the difference in quantum field statistics depending on the variance of geometry and topological phases of matter.

Although the above formalism is valid for any type of material system, here we focus on translationally invariant systems that do not mix electrons with different momenta. In this case, the current operator is diagonal in momentum space, i.e., $\hat{j}_p(t)=\sum_{\textbf{k}}\hat{j}^{p}_{\textbf{k}}(t)$, where $\textbf{k}$ is the Bloch wave vector of the lattice. Therefore, it can be shown that the correlation tensor $D$ grows linearly with the number of emitters $N$, as it can be rewritten as
\begin{equation}
\begin{aligned}
    D_{\mu,\nu}(t,t')&=\sum\limits_{\textbf{k}}\langle j^p_{\mu,\textbf{k}}(t) j^p_{\nu,\textbf{k}}(t')\rangle-\langle j^p_{\mu,\textbf{k}}(t)\rangle\langle j^p_{\nu,\textbf{k}}(t')\rangle\\&=\sum\limits_{\textbf{k}}\langle \delta j^p_{\mu,\textbf{k}}(t) \delta j^p_{\nu,\textbf{k}}(t')\rangle,
\end{aligned}\label{Eq:D_tensor_k}
\end{equation}
where $\delta j^p_{\mu,\textbf{k}}(t)=j^p_{\mu,\textbf{k}}(t)-\langle j^p_{\mu,\textbf{k}}(t)\rangle$ and thus, this tensor can be interpreted as a covariance tensor.
We note that in the case of the length gauge, instead of the current-current correlation function, we would obtain the dipole-dipole correlation function. Under the external field, electrons are able to move throughout the crystal structure and hence their coordinates might be correlated on such length scales, which results in a possible divergence of the dipole-dipole correlations. Thus, we choose the velocity gauge to avoid such issues. 

Although the correlation matter tensor has a complicated structure in the presence of strong laser fields, for a periodic laser field it can be evaluated using the Floquet formalism and Wick's theorem. For zero or negligible dephasing in the material, it reads 
\begin{equation}
    D_{\mu,\nu}(t,t')=\sum\limits_{\substack{\alpha\beta \\ \alpha'\beta'}}\sum\limits_{\textbf{k}}f^{0,\textbf{k}}_{\alpha\beta}\left(\delta_{\alpha'\beta'}-f^{0,\textbf{k}}_{\alpha'\beta'}\right)J^{\mu,\textbf{k}}_{\beta\alpha'}(t)J^{\nu,\textbf{k}}_{\beta'\alpha}(t'),\label{Eq:FloquetD}
\end{equation}
where the indexes $\alpha,\beta$ denote Floquet band numbers, $f^{0,\textbf{k}}_{\alpha\beta}$ is a matrix element of the occupation number operator in the absence of dephasing mechanisms in the solid-state system. It is determined by the distribution function $F^{0}_{\alpha\beta}=\delta_{\alpha\beta}-2f^{0}_{\alpha\beta}$, which connects the Keldysh correlation function with the retarded and the advanced Green functions through a space–time convolution. Finally, $J^{\mu,\textbf{k}}_{\alpha\beta}(t)$ is the current matrix element between the $\alpha$ and $\beta$ Floquet bands. See Appendix~\ref{app:FP} for the detailed discussion. 

In the absence of pumping, the matter system is in thermal equilibrium, so there are no off-diagonal elements in the distribution function and $f^{0,\textbf{k}}_{\alpha\beta}\rightarrow f_{m,\textbf{k}}\delta_{\alpha\beta}\delta_{m\alpha}$. Consequently, correlation matter tensor depends only on the relative time only as $D_{\mu,\nu}(t,t')=D_{\mu,\nu}(t-t')$ and hence
\begin{equation}
\begin{aligned}
        D^{(0)}_{\mu,\nu}(\tau)&=\sum\limits_{m\neq p,\textbf{k}}f_{m,\textbf{k}}(1-f_{p,\textbf{k}})\epsilon^2_{mp,{\textbf{k}}}r^{mp}_{\mu,\textbf{k}}r^{pm}_{\nu,\textbf{k}}e^{-i\epsilon_{pm,\textbf{k}}\tau},
\end{aligned}\label{Eq:zeroD}
\end{equation}
where the indexes $n,m$ denote the band numbers, $\epsilon_{m,\textbf{k}}$ is the energy of an electron in the band $m$ with momentum $k$, $\epsilon_{nm,\textbf{k}}\equiv\epsilon_{n,\textbf{k}}-\epsilon_{m,\textbf{k}}$, $f_{m,\textbf{k}}=(1+e^{\beta_{th}(\omega_{\textbf{k},m}-\mu_0)})^{-1}$ is the Fermi-Dirac distribution function, $\beta_{th} = 1/(k_BT_{th})$, $k_B$ is the
Boltzmann constant, $T_{th}$ is the absolute temperature of the thermal bath, $\mu_0$ is the chemical potential, $j^{mn}_{\mu,\textbf{k}}=-i\epsilon_{mn,\textbf{k}}\ r^{mn}_{\mu,\textbf{k}}$ is the matrix element of the paramagnetic current operator, $r^{mn}_{\mu,\textbf{k}}=\langle u^{m}_{\textbf{k}}|i\partial_{\mu}u^{n}_{\textbf{k}}\rangle$ the momentum-space non-Abelian Berry connection. The sum over momentum runs over the first Brillouin zone (FBZ).  

Using Eq.\eqref{Eq:zeroD} together with the relation between the susceptibility and the correlation matter tensor, one recovers the standard Kubo-formula, as expected by construction. Moreover, a comparison with the explicit expressions in~\cite{Cheng:2010oyg, Verma} shows that the correlation matter tensor is closely related to the quantum geometric tensor (QGT) as
\begin{equation}
    D^{(0)}_{\mu,\nu}(\tau)=-\frac{\partial^2}{\partial\tau^2}\mathcal{Q}_{\mu,\nu}(\tau).\label{Eq:DQ}
\end{equation}
This relation shows that the geometrical and topological properties of the system leave their imprint on the photonic signatures of the emitted light. As will be shown below, these correlations play a crucial role in the quantum properties of the radiated field. From a macroscopic point of view, this relation is natural: since the current and the polarization in the material are related by $\langle \hat{j}_{\mu}(t)\rangle=d\langle \hat{d}_{\mu}(t)\rangle/dt$, an analogous relation is expected between current–current and dipole–dipole fluctuations, which, for an instantaneous response, reduces to Eq.~\eqref{Eq:DQ}. See the Appendix~\ref{app:D3} for more details.

In HHG, however, the laser field is highly intense, so we cannot restrict ourselves Eq.\eqref{Eq:FloquetD}. Moreover, a solid-state system is generally coupled to an external thermal bath, which sets the temperature and, consequently, the equilibrium distribution of electrons over the bands. It is well-known \cite{ PhysRevLett.113.073901} that the main source of high-order harmonics is interband transitions, which are very sensitive to dephasing. Therefore, the dephasing and relaxation processes governed by the bath must be taken into account to describe the correlations in the material properly. Meanwhile, the periodic driving pushes the system away from thermal equilibrium towards Floquet-induced stationary states, which differ from the thermal ones. Once the system reaches such a new stationary state, interband transitions cease and HHG becomes inefficient. Hence, the problem is, in general, of non-equilibrium nature. A consistent way to account for both processes is to solve the corresponding Floquet kinetic equation~\cite{PhysRevA.92.062108}. To this end, one first solves the bare Floquet problem of the driven material and then uses the resulting Floquet eigenbasis to calculate the self-energy. This method yields the drive-induced change in relaxation and the new stationary electron distribution, from which all required correlation functions follow via the spectral functions. The explicit definition of the Floquet problem, the detailed derivation of the Floquet kinetic equation, and the correlation matter tensor within the Keldysh formalism of non-equilibrium Green functions are given in Appendixes~\ref{app:FP} and~\ref{app:FKE}, one can find an explicit definition of the Floquet problem, the detailed derivation of the Floquet kinetic equation as well as correlation matter tensor based on the Keldysh formalism of non-equilibrium Green functions. We do not reproduce these derivations here, since the main goal of this paper is to connect the correlation matter functions with the statistics of photons emitted in HHG and to show their correspondence. In the next section, it will become clear which particular components of the correlation matter tensor can be accessed through photon statistics; the calculation of these components of $D_{\mu,\nu}$ is discussed there.

Whereas previous HHG theories proceed with the direct solution of the corresponding Schrödinger equation for the wavefunction (the ULL equation for the density matrix in our case), we instead use the ULL equation to write the equations of motion for field statistics.

To obtain results that can be easily interpreted in the examples below, we implement the incident laser field with two bright coherent tones at $\pm\Omega$ as 
\begin{equation}
    A(t)=\begin{cases}
        A_0\sin(\Omega t), \ t\leq0\leq t_f \\
        0, \ \ \ \ \ \ \ \ \ \ \ \ \ \text{otherwise}& 
    \end{cases},
\end{equation}
where $A_0$ is the modulation amplitude, $\Omega$ is the modulation frequency and $t_f=N_{cyc}T$ is the total duration of the laser field and $N_{cyc}=10$ as ten cycles per one pulse. Another advantage of such a simple field is that it is easy to generate experimentally.

\section{Field statistics}
\label{sec:field_statistics}

\begin{figure}[t]
\centering
\includegraphics[width=0.45\textwidth]{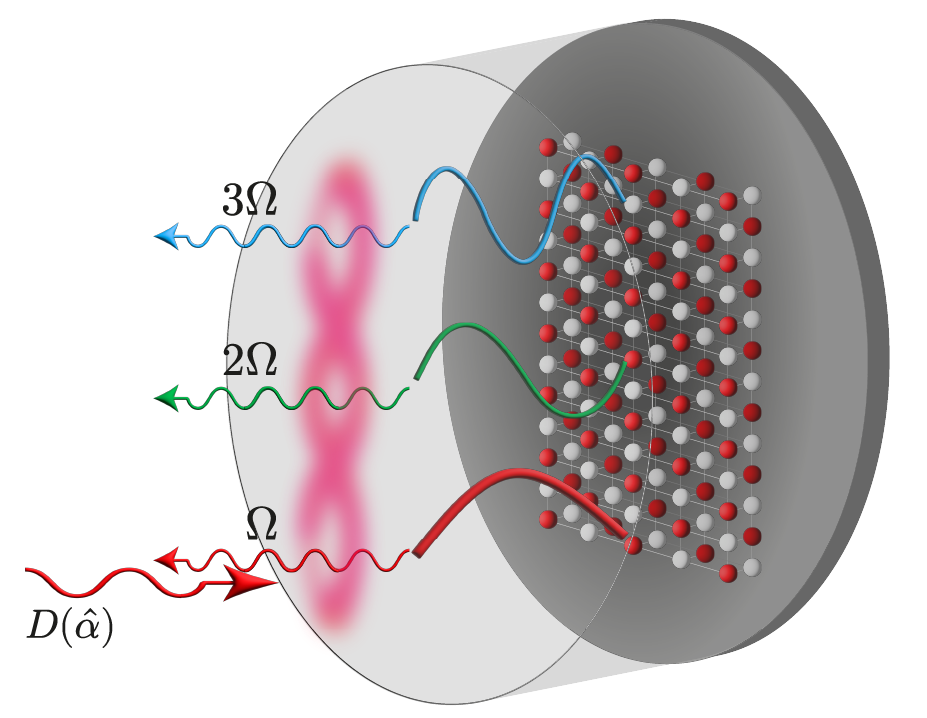}
\caption{The crystal structure is placed inside a one-sided cavity. An external multimode coherent-state laser enters the cavity and interacts with the material system. The currents generated inside the system emit light and populate the intracavity modes. As a result, each mode starts radiating outside the cavity.}\label{fig:cavity}
\end{figure}

As mentioned in the previous section, no strict restriction on the photonic surroundings was imposed to derive the aforementioned formalism except for weak light-matter coupling. In this part and further, we restrict our consideration to a particular case of a single-sided multi-mode cavity, which is a Fabry-Pérot cavity with an embedded back mirror. The laser enters the cavity, excites the matter system embedded in the cavity, which results in emitting light inside and subsequent emission outside the cavity. To implement the open-type photonic surroundings, we phenomenologically introduce cavity losses via an effective decay rate $\gamma$ for the fundamental mode (see Fig.~\ref{fig:cavity}). In the case of radiation in a cavity, instead of determining the radiation mode via the wave vector $\textbf{q}_{\mu}$ we can do it by introducing a cavity mode number $m_{\mu}$ and the frequency of radiation $\omega_{m}$. For simplicity, we consider a cavity whose relevant resonances are tuned to harmonic frequencies such as $\omega_m=m\Omega$ and $m\in \mathbb{N}$. Finally, we assume that the cavity has only $M$ modes for each polarization, whereas the high quality factor is considered for the fundamental mode, i.e. $Q_c=\Omega/2\gamma\gg1$. Because the decay rate increases linearly with the mode number, we assume constant quality factor for any mode.  

We start from the boundary condition relating each of the far field amplitudes outside the cavity to the intracavity field. The outgoing field can be find as a sum over all intracavity channels as
\begin{equation}
    a^{\mu}_{out}(t)=\sum\limits_{m}\sqrt{2\gamma_m}a_{\mu,m}(t)-a^{\mu}_{in}(t).\label{Eq:outin}
\end{equation}
As a result, interference terms between the input and the cavity field may contribute to the observed moments outside the cavity. We note that initially in the rotated frame there is no incoming field, so $\langle a^{\mu}_{in}(t)\rangle=0$, because the laser field is taken into account as a classical driving force for a material system.

Finally, we can write down the equation of motion for the intracavity field operators. Taking into account the explicit value of the average current in the material system, Eq.~\eqref{Eq:current}, the field amplitude operator inside the cavity is

\begin{equation}
\begin{aligned}
    \frac{d\langle a_{\mu,m}(t)\rangle}{dt}=&-(\gamma_m+i\omega_m) \langle a_{\mu,m}(t)\rangle-\frac{ig_0}{\sqrt{\omega_m}}\langle j_{\mu}(t)\rangle_0 \\&-\frac{ig_0}{\sqrt{\omega_m}}\int\limits_{-\infty}^{t} dt'\sigma_{\mu,\nu}(t,t')\langle E^{(\nu)}_Q(t')\rangle,
\end{aligned}\label{Eq:amplitude}
\end{equation}
where $\sigma_{\mu,\nu}(t,t')$ is the dynamical conductivity tensor and $E^{(\nu)}_Q=-\partial A^{(\nu)}_Q/\partial t$ is the electric field operator. The obtained equation for the field amplitude has a clear physical meaning, namely, the first line corresponds to the classical generation of the field via classically induced paramagnetic current and leakage outside the cavity, whereas the second line corresponds to the back-action phenomenon, which is a retarded effect of the emitted fields via nonlinear conductivity, which in turn induces extra currents in the material system as a correction to the classical paramagnetic one. As a result, it is particularly convenient to solve this equation using the following ansatz
\begin{equation}
    \langle a_{\mu,m}(t)\rangle=a^{(cl)}_{\mu,m}(t)+a^{(q)}_{\mu,m}(t),\label{Eq:exp_amp}
\end{equation}
where the first classical term represents the semiclassical HHG answer and grows as $\sim g_0N$, while the second term is a quantum correction to the classical term and scales {$\sim g^{3}_0N^{2}$. Despite the fact that we associate the second term with the quantum correction, it has a purely classical meaning in the sense that its occurrence can be anticipated from the point of view of classical electrodynamics. Moreover, the Lindbladian part of the ULL equation has no influence on the dynamics of the field amplitude, and hence the total dynamics is governed by the unitary rotation in the ULL equation. Therefore, the quantum meaning of this term should be understood as that of only the next term in the expansion of the perturbative series in terms of the weak coupling strength. This fact is crucial in understanding the quantum features of light inside or outside the cavity. Despite its apparent smallness, we will keep this quantum term to produce accurate calculations of the emitted field statistics further.

Turning to two-operator statistics, it is reasonable to start from the intracavity mode population $n_{\mu,m}=\langle a^{\dagger}_{\mu,m}a_{\mu,m}\rangle$. The corresponding equation of motion describes the energy change of the field inside the cavity over time. (see Appendix~\ref{app:B}). However, the most interesting characteristic is a variation $\delta n_{\mu,m}(t)=n_{\mu,m}(t)-n^{(coh)}_{\mu,m}(t)$ that describes the deviations of the average population from the coherent population $n^{(coh)}_{\mu,m}=\langle a^{\dagger}_{\mu,m}\rangle\langle a_{\mu,m}\rangle$. These deviations are responsible for departures from the multi-mode coherent state, as well as for non-trivial statistical features of the emitted light. After some algebra, the equation for the intracavity population deviation can be written as
\begin{equation}
    \begin{aligned}
        n_{\mu,m}(t)&=n^{(coh)}_{\mu,m}(t)+\overline{n}_{th}+\int\limits_{-\infty}^{t}dt'S_{\mu,m}(t')e^{-2\gamma_m(t-t')}
\end{aligned}\label{Eq:population}.
\end{equation}
The first term in Eq.~\eqref{Eq:population} is a coherent contribution. The second one corresponds to the population of the thermal bath, which obeys the standard Bose-Einstein statistics. However, because the characteristic frequencies of HHG process are very high, even at room temperature one can neglect this population. The third contribution is the response from the Lindbladian term in Eq.~\eqref{Eq:ull}. This term is of a purely quantum nature, as it arises from the commutation relation between photonic operators and represents the spectral noise function. It can be written as
\begin{equation}
\begin{aligned}
    S_{\mu,m}(t)&=\frac{g^2_0}{\omega_m}\int\limits_{-\infty}^{\infty}d\tau D_{\mu,\mu}(t+\tau,t)e^{-i\omega\tau}.
\end{aligned}\label{Eq:quan_pop}
\end{equation}
From the definition of the correlation matter tensor Eq.~\eqref{Eq:D_tensor_k} we can deduce that the quantum correction $\delta n_{\mu,m}$ is nonnegative due to the similarity of the expression with the spectral density. It can be noticed, that the quantum contribution grows as $\sim g^2_0N$, whereas the coherent part $\sim g^2_0N^2(1+g^2_0N)$, where the first part corresponds to the semiclassical response and the second to the interference between the classical field and quantum correction for field amplitudes Eq.~\eqref{Eq:exp_amp}. From such behaviour, it is obvious that for extremely large systems the quantum noise term decays to zero, which reflects the classical limit, as expected. Nevertheless, only the quantum noise induced population can break the multi-mode coherent nature of the emitted fields and hence is responsible for quantum light generation. The fact that quantum features arise due to the nonlinear nature of the correlation matter tensor leads to the idea that all light statistics are highly sensitive to quantum geometry of the material system, correlations and interactions of electrons within it. Thus, by exploring emitted field statistics, one can study the transport properties of material systems and hence open a new avenue for study.

For our purposes in this survey, we also need to find the average squared field for each mode, namely $\langle a^2_{\mu,m}(t)\rangle$. As before, of particular interest is the equation for deviations from the coherent field $\delta a^2_{\mu,m}(t)=\langle a^2_{\mu,m}(t)\rangle-\langle a_{\mu,m}(t)\rangle^2$.  Thus, by performing the same algebra, we obtain
\begin{equation}
    \begin{aligned}
        \langle a^2_{\mu,m}(t)\rangle&=\langle a_{\mu,m}(t)\rangle^2\\&-\int\limits_{-\infty}^{t}dt'\Xi^{mm}_{\mu\mu}(t')   e^{-2(\gamma_m+i\omega_m)(t-t')},\end{aligned}\label{Eq:a_sq}
\end{equation}
where 
\begin{equation}
\begin{aligned}
\Xi^{p_1p_2}_{\mu\nu}(t)&=\frac{2g^2_0}{\sqrt{\omega_{p_1}\omega_{p_2}}}\\&\times\Bigg(\frac{i\langle j_{d,\mu\nu}(t)\rangle}{2}+\int\limits_{0}^{\infty} d\tau D_{\mu,\nu}(t+\tau,t) e^{-i\omega_{p_2}\tau}\Bigg),\label{Eq:xi_m}
\end{aligned}
\end{equation}
where $\Xi^{p_1p_2}_{\mu\nu}(t)$ is a generalized noise correlation tensor. As can be seen, the spectral noise function is proportional to the real part of the generalized noise correlation function, i.e. $\Xi^{mm}_{\mu\mu}(t)=S_{\mu,m}(t)+I_{\mu,m}(t)$, where $I$ defines the reactive response of the material. As before, the first term in Eq.~\eqref{Eq:a_sq} corresponds to the coherent fields, whereas the second one arises due to the quantum noise, which in turn is responsible for  the generation of quantum light states. The scaling of the expression is the same as for population in Eq.~\eqref{Eq:population}.

By proceeding in the same way, it is possible to consider the higher-order statistics of the fields. However, this is outside the scope of the present paper.

Turning to the statistical properties of the emitted field, one should distinguish between two cases: the intracavity field and the far-field emission. The former characterizes the light inside the cavity, whereas the latter characterizes the field leaving the cavity. In the present study, we concentrate on far-field statistics because of their practical importance. However, since the outgoing and intracavity fields are directly related through Eq.\eqref{Eq:outin}, we first need to study the intracavity photon correlations. A detailed discussion is given in Appendix~\ref{app:C}.

\section{Far-field photon properties}
\label{sec:far_field}

The developed fully quantum-mechanical treatment of the HHG process makes it possible to determine not only the emission spectrum — the standard observable in conventional HHG analyses — but also additional properties of the emitted field. In particular, since this method provides access to the photon number distribution in each mode, one can evaluate the photon statistics and thus explore signatures of non-classical behavior in HHG. Below, we show how to compute the second-order correlation function $g_2(\tau)$~\cite{Scully_Zubairy_1997}, which can be measured via photon counting~\cite{PhysRevLett.110.173602}, the squeezing of the single-mode photonic state, which can be found experimentally via homodyne detection method~\cite{Breitenbach}.

To calculate any type of far-field statistics, one needs the two-time correlation functions of the output operator $a^{\mu}_{out}$. These can be obtained with the help of the quantum regression theorem~\cite{Zoller:1997su, Swain_1981,PhysRevA.99.033816, PhysRevA.109.052207}, which states that, whenever the master equation yields linear equations of motion for the mean values, the same equations determine multi-time correlation functions with modified initial conditions. A detailed discussion and derivation of all necessary two-time correlation functions are given in Appendix~\ref{app:C}.

As a first example, we consider the intensity spectrum of the outgoing $\mu$-polarized field at frequency $\omega$ 
\begin{equation}
    S_{\mu}(\omega)=\int\limits_{-\infty}^{\infty}d\tau e^{i\omega \tau}\overline{\langle a^{\mu^{\dagger}}_{out}(t)a^{\mu}_{out}(t+\tau)\rangle}^t,\label{Eq:intensity}
\end{equation}
where $\overline{\langle A(t)B(t+\tau)\rangle}^t$ denotes the average over one period of the laser field. After some algebra, the spectrum can be decomposed into coherent and incoherent parts:
\begin{equation}
    \begin{aligned}
        &S_\mu(\omega)=S_{coh}(\omega)+S_{incoh}(\omega),\\
        &S_{coh}(\omega)=|F(\omega)|^2|\langle j_{\mu}(\omega)\rangle|^2\sim N^2
        ,\\
        &S_{incoh}(\omega)=|F(\omega)|^2D_{\mu\mu}(-\omega,\omega)\sim N,
    \end{aligned}\label{Eq:anal_intens}
\end{equation}
where $D_{\mu,\nu}(\omega_1,\omega_2)=\int dt\int dt'e^{i\omega_1t}e^{i\omega_2t'}D_{\mu,\nu}(t,t')$, $F$ denotes the multi-mode transfer function $F(\omega)\equiv\sum_mg_0G^m_c(\omega)/\sqrt{Q^m_c}$ and $G^m_c(\omega)=(\omega-\omega_m+i\gamma_m)^{-1}$ is the single-pole propagator of cavity mode $m$. Effectively, this function acts as a filter that enhances the resonant transitions and suppresses non-resonant scattering. To leading order in $g_0$, the cavity enters only through $F$. As can be seen, the coherent part of the spectrum grows as $N^2$, whereas the incoherent part grows only as $N$. Consequently, in macroscopic systems only the coherent contribution is relevant, whereas fundamentally different behaviour can arise in finite mesoscopic structures. We therefore anticipate that the statistical properties of the outgoing field are very sensitive to the number of atoms in the structure. Moreover, because the incoherent part is determined by current–current fluctuations, it is deeply connected to the geometry of the material through the quantum geometric tensor. 

One can check that in the absence of the laser field and at zero temperature of the photon bath, the cavity emits no real, measurable photons. Indeed, in this case the average current vanishes, $\langle j_{\mu}(t)\rangle\equiv0$. The current–current correlations, however, are nonzero and given by Eq.\eqref{Eq:zeroD}. Nevertheless, we can notice that $S_{incoh}(\omega)\sim D_{\mu\mu}(-\omega,\omega)\sim\delta(\omega+\epsilon_{pm,\textbf{k}})$, where by construction in Eq.\eqref{Eq:zeroD} $\epsilon_{pm,\textbf{k}}>0$. As a result, the delta-function condition is never satisfied, so the matter-induced noise does not contribute to the emission. Hence, no photons are emitted in the absence of pumping and thermal fluctuations, as expected.

The spectral power can be viewed as a first-order correlation function and therefore does not reveal all the statistical properties of the underlying photon distribution. To access higher-order correlations of the generated light, one calculates the second-order correlation function. The value $g_2(\omega)_{\mu}=1$ corresponds to Poissonian statistics, which a coherent field possesses, whereas for $g_2(\omega)_{\mu}>1$ the photon statistics are called super-Poissonian while for $g_2(\omega)_{\mu}<1$ they are called sub-Poissonian, referring to a broader and a narrower distribution than a Poissonian distribution, respectively. We note that super-Poissonian statistics is not necessarily a quantum feature, since classical mixtures of coherent states, thermal states, and non-classical states can all exhibit it. In contrast, only a non-classical state can produce sub-Poissonian statistics. The choice of the particular correlation function is discussed in detail in Appendix~\ref{app:H}. In this study, we consider only a photon counting scheme for the fixed mode $(\omega,\mu)$, for which the correlation function reads
\begin{equation}
    g_2(\omega)_{\mu}=\frac{\langle a^{\mu^{\dagger}}_{out}(\omega)a^{\mu^{\dagger}}_{out}(\omega)a^{\mu}_{out}(\omega)a^{\mu}_{out}(\omega)\rangle}{\left(\langle a^{\mu^{\dagger}}_{out}(\omega)a^{\mu}_{out}(\omega)\rangle\right)^2}.\label{Eq:g2w}
\end{equation}
After some algebra, we can obtain 
\begin{equation}
    \begin{aligned}
g_2(\omega)_{\mu}&=2-\frac{1}
{(s(\omega)+1)^2}\\&+\frac{\left|1+\frac{D_{\mu\mu}(\omega,\omega)}{\langle j_{\mu}(\omega)\rangle^2}+\frac{\omega\sigma_{\mu\mu}(\omega,\omega)\rangle}{\langle j_{\mu}(\omega)\rangle^2}\right|^2-1}{(s(\omega)+1)^2}  ,\end{aligned}\label{Eq:g2wans}
\end{equation}    
where $s(\omega)=S_{incoh}(\omega)/S_{coh}(\omega)\sim 1/N$ and $\sigma_{\mu\nu}(\omega_1,\omega_2)$ is a Fourier transform of the conductivity tensor. The first line in the above formula originates from the normal correlation function. As can be seen, emission always makes it necessary to obtain super-Poissonian statistics for the mesoscopic system size. The second line corresponds to the anomalous contribution, whose sign is not fixed. Consequently, the anomalous correlation can drive the system towards sub-Poissonian or super-thermal statistics. In contrast, in a macroscopic system all incoherent corrections are negligible, and the statistics approach those of coherent emission.

The last quantity characterizing single-mode statistics is the degree of far-field squeezing, which is an inherently non-classical feature of light. To estimate it, we calculate the noise power spectrum of the quadrature $\hat{X}_{\theta}(t)=(\hat{a}^{\mu}_{out}(t)e^{-i\theta} +\hat{a}^{\mu^\dagger}_{out}(t)e^{i\theta} )/2$ rotated by an angle $\theta$
\begin{equation}
    \Delta\hat{X}_{\mu,\omega}(\theta) = \frac{1}{4}+\int\limits_{-\infty}^{\infty}d\tau e^{i\omega\tau}\overline{\langle :\hat{X}_{\theta}(t),\hat{X}_{\theta}(t+\tau):\rangle}^t,
\end{equation}
where $\langle:..:\rangle$ denotes normal ordering of operators. The degree of squeezing of the mode in decibels is given by
\begin{equation}
r_{\mu}(\omega)=-10\log_{10}\left(4\min_{\theta\in [0,\pi)}\Delta \hat{X}_{\mu,\omega}(\theta)\right),\label{Eq:degree_sq}
\end{equation}
where the minimum taken over the quadrature angle $\theta$. Classical coherent light is not squeezed, i.e. $r_{\mu}(\omega)=0$ for all modes and polarizations of coherent light. With the help of the quantum regression theorem and after optimizing over all angles $\theta$, we obtain the noise power spectrum at the optimal angle 
\begin{equation}
\begin{aligned}
    \Delta\hat{X}_{\mu,\omega}(\theta_{opt})=\frac{1}{4}&+\frac{1}{2}\left(S_{sym}(\omega)-|M(\omega)|\right),
\end{aligned}\label{Eq:min_quad}
\end{equation}
where $S_{sym}(\omega)=(S_{incoh}(\omega)+S_{incoh}(-\omega))/2$ is the symmetrized incoherent power spectral density and $M(\omega)$ is the anomalous spectral density, which is given by 
\begin{equation}
    M(\omega)=F(\omega)F(-\omega)\Big[D_{\mu\mu}(-\omega,\omega)-\omega\sigma_{\mu\mu}(\omega,-\omega)\Big].\label{Eq:M_anom}
\end{equation} 
Therefore, the outgoing radiation is squeezed if $\Delta\hat{X}_{\mu,\omega}(\theta_{opt})<1/4$, or equivalently, $S_{sym}(\omega)<|M(\omega)|$; that is, the anomalous spectral density must exceed the symmetrized power spectral density. Notably, because of the properties of coherent states, this condition does not depend on the coherent field at all; it is determined solely by the noise correlation function and the conductivity tensor. This supports our statement that the noise correlations, which are determined by current–current fluctuations in the solid, are responsible for quantum light generation via the Kerr nonlinearity. The conductivity term, however, provides an additional source of squeezing: by construction, it acts as a natural parametric amplifier.

Finally, we study the two-mode correlations of the outgoing fields. To do so, we extend the definition of the single-mode second-order correlation function in Eq.\eqref{Eq:g2w} and consider the correlation function between modes $\omega_1$ and $\omega_2$ as
\begin{equation}
    g_2(\omega_1,\omega_2)_{\mu}=\frac{\langle a^{\mu^{\dagger}}_{out}(\omega_1)a^{\mu^{\dagger}}_{out}(\omega_2)a^{\mu}_{out}(\omega_2)a^{\mu}_{out}(\omega_1)\rangle}{\langle a^{\mu^{\dagger}}_{out}(\omega_1)a^{\mu}_{out}(\omega_1)\rangle\langle a^{\mu^{\dagger}}_{out}(\omega_2)a^{\mu}_{out}(\omega_2)\rangle}.\label{Eq:g2w1w2}
\end{equation}
Simplifying it in the same way, we obtain
\begin{equation}
    \begin{aligned}
g_2(\omega_1,&\omega_2)_{\mu}=1+\frac{\left|1+\frac{D_{\mu\mu}(-\omega_1,\omega_2)}{\langle j_{\mu}(-\omega_1)\rangle\langle j_{\mu}(\omega_2)\rangle}\right|^2-1}
{(s(\omega_2)+1)(s(\omega_1)+1)}\\&+\frac{\left|1+\frac{D_{\mu\mu}(\omega_1,\omega_2)}{\langle j_{\mu}(\omega_1)\rangle\langle j_{\mu}(\omega_2)\rangle}+\frac{\omega_1\sigma_{\mu\mu}(\omega_2,\omega_1)\rangle}{\langle j_{\mu}(\omega_1)\rangle\langle j_{\mu}(\omega_2)\rangle}\right|^2-1}{(s(\omega_2)+1)(s(\omega_1)+1)} .\end{aligned}\label{Eq:g2w1w2_ans}
\end{equation}    
The same discussion of single-mode $g_2$ applies equally to the two-mode case. Again, the main source of non-classical statistics is the anomalous part, which becomes significant in the mesoscopic regime.

\section{One-dimensional Su-Schrieffer-Heeger}
\label{sec:1d_ssh}

As an example of our general quantum HHG formalism we choose the one-dimensional Su–Schrieffer–Heeger (1D SSH) model, the simplest canonical model of a topological insulator in one dimension~\cite{PhysRevLett.42.1698}. Its topological properties in the static case have been studied in detail~\cite{PhysRevB.100.075437, Meier}. Moreover, its periodically driven version has been used to analyze Floquet phases~\cite{10.1038/s42005-024-01908}, and the model has already been applied to HHG~\cite{PhysRevA.110.033113, PhysRevB.99.195428, lange2026edgestates}. 

In the context of HHG, the model can be viewed as a chain of $2N$ atoms with a bipartite lattice structure, whose hopping amplitudes are modulated in time by the driving field, and with a single photonic polarization along the chain, as depicted in Fig.~\ref{fig:1dssh}(a). The equilibrium atomic positions are taken as
\begin{equation}
  z_i = \left(i - \frac{N+1}{2}\right)a - (-1)^i \delta, 
  \qquad i = 1, 2, \ldots, N,
  \label{eq:positions}
\end{equation}
where $a$ is the mean nearest-neighbor spacing and $\delta$ (with $0 < |\delta| < a/2$) is a small, nonzero displacement that encodes the Peierls dimerization. Within a nearest-neighbor tight-binding description, the hopping amplitude between adjacent atoms decays exponentially with their separation, which yields the intracell and intercell hopping parameters
\begin{equation}
\begin{aligned}
  J_1 &\equiv -e^{-(z_{2m} - z_{2m-1})} = -e^{-(a - 2\delta)},\\
  J_2 &\equiv -e^{-(z_{2m+1} - z_{2m})} = -e^{-(a + 2\delta)}.  
\end{aligned}\label{eq:hopping_w}  
\end{equation}
In the static case the Hamiltonian can be written as 
\begin{equation}
    \begin{aligned} \hat{H}_{SSH}=\sum\limits_{j=1}^{N/2}J_1\hat{c}^{\dagger}_{A,j}\hat{c}_{B,j} +\sum\limits_{j=1}^{N/2-1}J_2\hat{c}^{\dagger}_{A,j+1}\hat{c}_{B,j}+\text{h.c.},
    \end{aligned}
\end{equation}
where $\hat{c}^{\dagger}_{A(B),j} (\hat{c}_{A(B),j})$ is the creation (annihilation) operator for the sublattice $A(B)$ at site $j$. We assume that the $A$ and $B$ sites have the same on-site energy. In such case, the open system in trivial insulator phase $(\delta>0)$ corresponds to the system with periodic boundary condition, because there are no edge states and all atoms are equivalent. For periodic boundary conditions, we can define the lattice operators in momentum space and rewrite the Hamiltonian $\hat{H}_{SSH}=\sum_k\Psi^{\dagger}_kH_{SSH}(k)\Psi_k$, where 
\begin{equation}
    H_{SSH}(k)=\begin{pmatrix}
0 & J_1+J_2e^{2ika} \\
J_1+J_2e^{-2ika} & 0
\end{pmatrix},
\end{equation}
and $\Psi_k=(\psi_A(k),\psi_B(k))$ and $k$ is an electron momentum. This model consists of two bands with dispersion $\epsilon_{\pm}(k)=\pm\sqrt{J^2_1+J^2_2+2J_1J_2\cos(2ka)}$ and band gap $\Delta=2|J_1-J_2|$. The topology of the model can be characterized by the Zak phase, which is obtained by integrating the Berry connection $r^{--}_{\mu,k}$ over the FBZ~\cite{Mondal}:

\begin{figure}[t]
\centering
\includegraphics[width=0.48\textwidth]{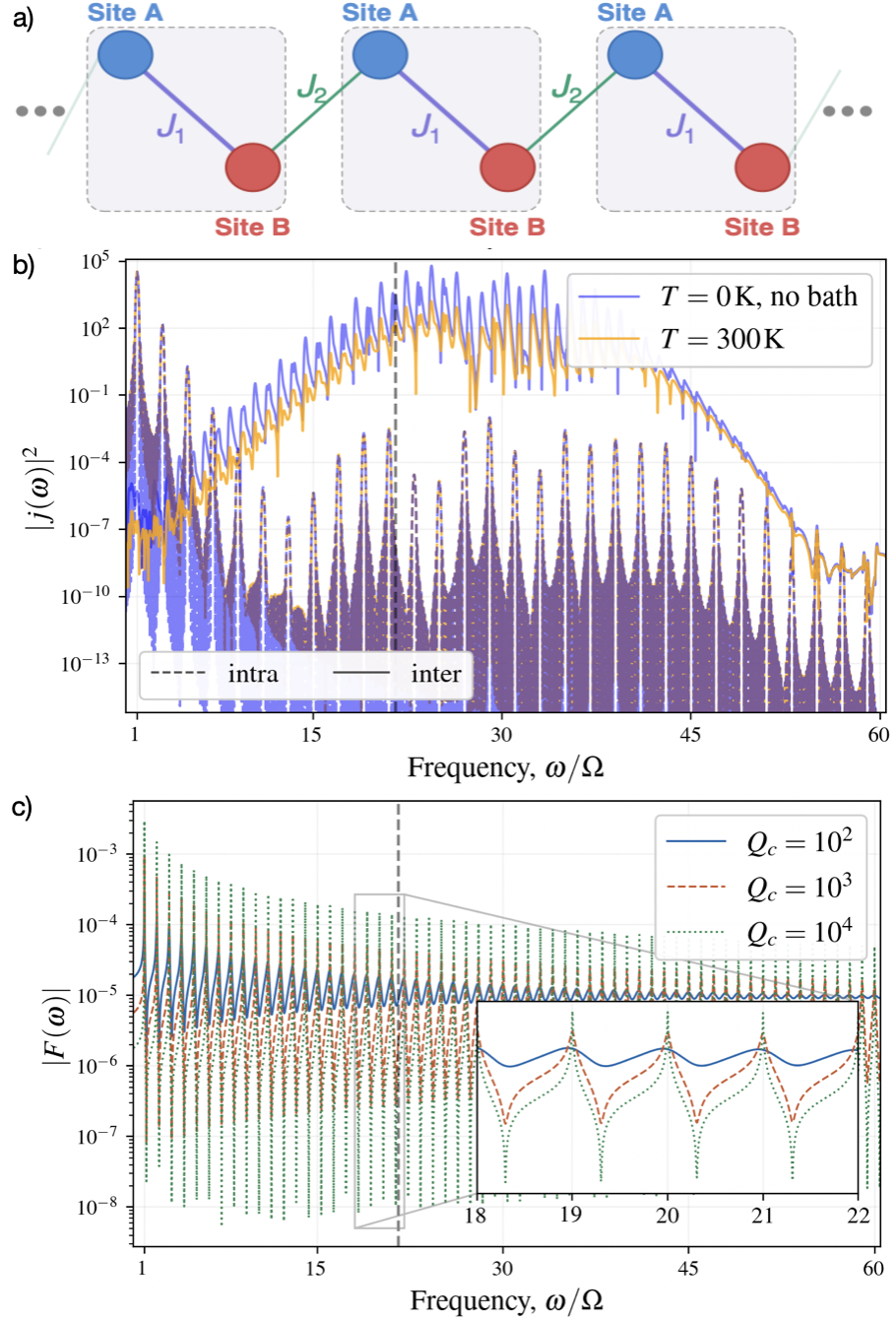}
\caption{a) The atomic structure of the one-dimensional model is presented; b) Squared absolute value of the current as a function of the frequency in the case of the absence of the bath (blue color) and in the presence of the bath (brown color). Solid (dashed) line corresponds to interband (intraband) transitions; c) absolute value of the filter function $F$ as a function of frequency for different cavity quality factors. The inserted plot is the zoomed version of the main. Black vertical dashed line determines band gap of the chain.}\label{fig:1dssh}
\end{figure}

\begin{equation}
\gamma_0
=
\int_{-\pi/2a}^{\pi/2a} r^{--}_{\mu,k}\, dk
=
\pi\Theta(J_2/J_1-1),
\end{equation}
and hence nontrivial topology occurs when $|J_2|>|J_1|$ which equivalent to $\delta<0$. However, since we consider periodic boundary conditions, the SSH chain has no edges and is therefore symmetric under the transformation $\delta\rightarrow-\delta$ ; hence, we fix $\delta>0$  in what follows. Under a periodic drive, the intra(inter)cell hopping becomes time-dependent and can be obtained via the Peierls substitution:
\begin{equation}
    J_{1(2)}(t)=J_{1(2)}e^{-iA(t)(a\mp2\delta)}.
\end{equation}

Before presenting the numerical results, we briefly discuss the physical range of parameters for the chosen model. Polyacetylene is a typical realization of the 1D SSH model~\cite{Grant, Meier, PhysRevB.99.195428} with $a=2\ \text{a.u.}$, $\delta=0.075a$ and $N=800$. These parameters correspond to a band gap of $\Delta=4.49eV$. The chain is pumped by the coherent infrared laser with $\lambda=6.1\mu m$ or equivalently $\Omega=0.0075\ \text{a.u.}$ during $N_{cyc}=10$ pulses. The magnitude of the vector potential is $A_0=0.2\ \text{a.u.}$. As a result, we obtain an $85$nm chain compared with a $102$nm wavelength for the 60th harmonic. This ensures that the validity of the dipole approximation. Finally, we assume weak light-matter coupling inside the cavity if the order $g_0N^2=1$. This condition guarantees that quantum coherent corrections are much smaller than the semiclassical response. We also neglect in-plane scattering and assume efficient in-phase transverse emission; the analysis of in-plane radiation effects is beyond the scope of this paper and is left for future research.

\begin{figure}[t]
\centering
\includegraphics[width=0.48\textwidth]{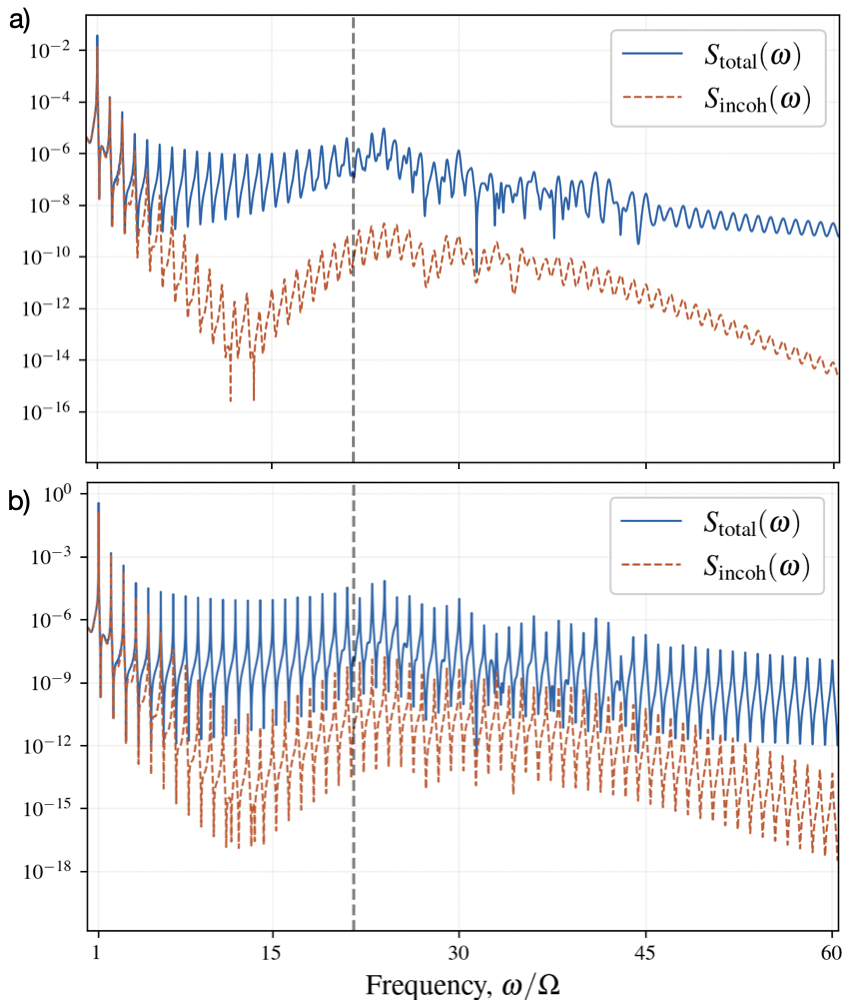}
\caption{Far-field emission spectrum is plotted for a) $Q_c=100$ and b) $Q_c=1000$. Blue color corresponds to the total spectrum, whereas brown color is related to incoherent part of the spectrum. Black vertical dashed line determines band gap of the chain.}\label{fig:spectrum}
\end{figure}

Although semiclassical HHG in SSH chains has already been studied in~\cite{PhysRevB.99.195428}, to the best of our knowledge, no work has yet addressed the quantum properties of the light emitted in this system.

We start our analysis with the spectrum of the average current induced in the chain by the laser field. In Fig.~\ref{fig:1dssh}(b), we explicitly separate the contributions of intraband (dashed lines) and interband (solid lines) transitions. See Appendix~\ref{app:D1} for more details. The first few harmonics are narrow and sharp and are formed predominantly by intraband transitions, whereas higher harmonics originate from interband transitions. This is consistent with the general features of HHG~\cite{PhysRevLett.113.073901}, and the HHG plateau is also observed. Moreover, this pattern holds both with (orange) and without (blue) the interaction with the thermal bath.

Here and below, we model the interaction of the electronic system with open space as an ohmic boson bath. By solving the corresponding kinetic equation (see Appendix~\ref{app:FKE} for details),, we obtain a dephasing mechanism that weakens the high-harmonic response by destroying the coherence of interband transitions. This is one of the main reasons why extremely short pulses are used in practice to maximize the efficiency of HHG. According to Eq.\eqref{Eq:anal_intens}, this plot also represents the semiclassical emission spectrum, which agrees well with~\cite{PhysRevB.99.195428}, except that a $\sin^2$ envelope was used there, whereas we use a pure sine wave.

Next, according to Eq.\eqref{Eq:anal_intens}, to leading order in $g_0$ the influence of the cavity on the emission spectrum factorizes from the matter correlation functions and is captured by the function $F(\omega)$.  It is therefore natural to examine its properties in order to understand how emission into the cavity, and from there into free space, differs from direct emission into free space. In Fig.~\ref{fig:1dssh}(c), we plot $F(\omega)$ for different cavity quality factors. This function acts as a filter: it enhances scattering at resonant harmonics and suppresses it at others. The higher the quality factor, the stronger the resonant enhancement. This is best seen in the zoomed-in inset, where $F(\omega)$ is smooth and broad for low quality factors and sharper and narrower for high ones. 

Taking the above into account, we plot the far-field emission spectrum,  Eq.\eqref{Eq:anal_intens} in Fig.~\ref{fig:spectrum}, where the first and the second panels correspond to the lower $Q_c=100$ and higher $Q_c=1000$ quality factors, respectively. The incoherent part is much smaller than the semiclassical response, except at a few low harmonics, where both are of the same order. This is a consequence of the mesoscopic regime, which is intermediate between single-atom emission, where both contributions are equally important, and macroscopic systems, where the incoherent part is negligible because it scales linearly $\sim N$, whereas the semiclassical radiation scales quadratically $\sim N^2$. Comparing the total emission from the cavity with the bare semiclassical response in Fig.~\ref{fig:1dssh}(b), we see that the cavity indeed acts as a filter that enhances resonant scattering and suppresses non-resonant contributions. Increasing the cavity quality factor strengthens this effect and yields more intense emission of the resonant modes.

\begin{figure}[t]
\centering
\includegraphics[width=0.48\textwidth]{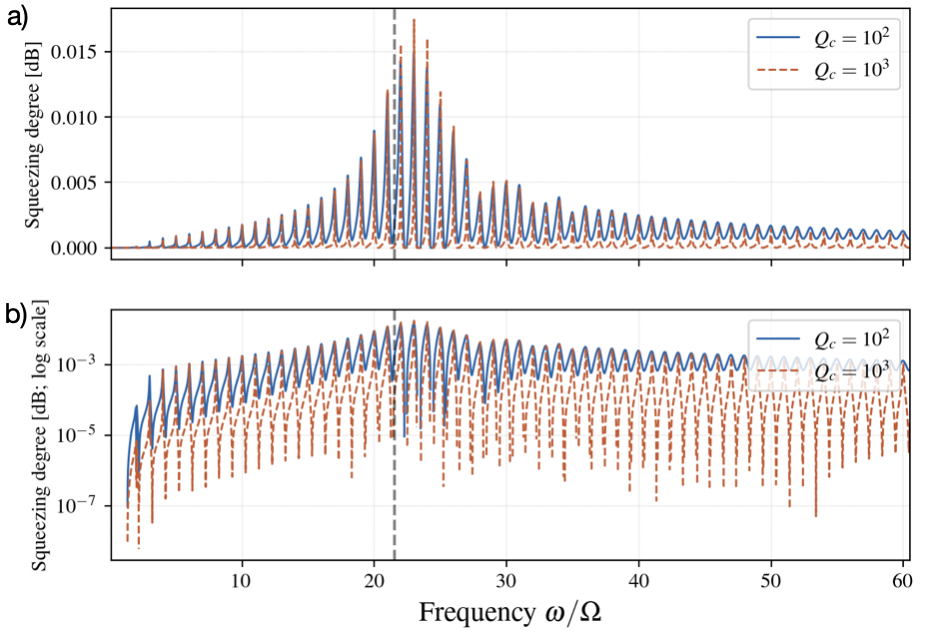}
\caption{Degree of single-mode squeezing calculated in dB in a) linear scale and in b) log scale for lower (blue color) and higher (orange color) cavity quality factor. Black vertical dashed line determines band gap of the chain.}\label{fig:squeezing}
\end{figure}

To better characterize and observe the quantum nature of the emitted field, we can calculate single-mode squeezing according to Eq.\eqref{Eq:degree_sq}, which is plotted in Fig.~\ref{fig:squeezing} on a linear scale in a) and on a logarithmic scale in b). Every mode is squeezed, and this squeezing is achieved with only a single coherent laser field. The squeezing profile follows that of the spectrum, mainly because of the filter function $F$. The most strongly squeezed modes are those that form the onset of the HHG plateau, as follows from their position relative to the band gap. As before, cavities with higher quality factors squeeze the outgoing field more strongly.

Although the obtained degree of squeezing is comparable to the experimental uncertainty reported in~\cite{Safavi-Naeini}, we believe that high-quality cavities can enhance the squeezing and make it detectable. According to Eq.\eqref{Eq:M_anom}, the squeezing is determined by current–current fluctuations and the longitudinal part of the conductivity tensor. This means that quantum geometry plays a crucial role in the formation of the quantum signatures of light emitted in HHG.

\begin{figure}[t]
\centering
\includegraphics[width=0.45\textwidth]{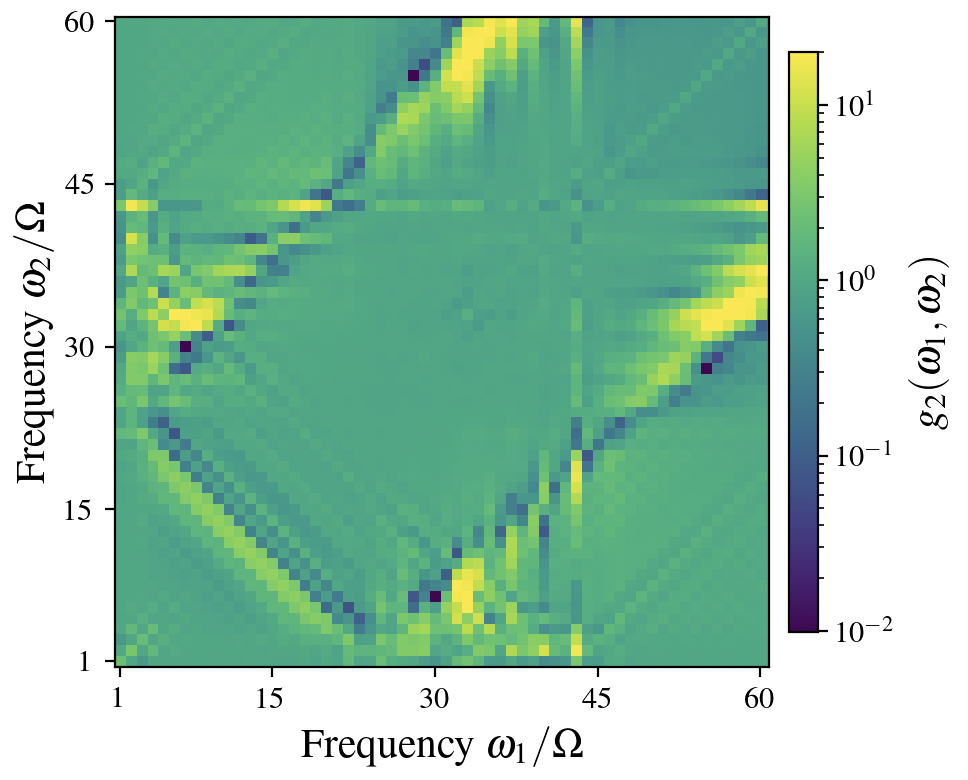}
\caption{Two-mode correlation function is presented. The diagonal of the colormap corresponds to the second order correlation function of the single mode.}\label{fig:two-mode}
\end{figure}

Finally, we calculate the correlation function between different modes according to Eq.\eqref{Eq:g2w1w2_ans} and show it in Fig.~\ref{fig:two-mode}. Most modes are uncorrelated. However, some pairs exhibit strong bunching or antibunching; that is, the probability of detecting both photons simultaneously is strongly enhanced or suppressed. This reveals nontrivial correlations between different modes induced by current–current fluctuations. Some of the high-order modes are even super-bunched. This result should be treated with caution, because, strictly speaking, at such high harmonic orders, terms beyond the dipole approximation may become relevant~\cite{Gorlach2020}. Nevertheless, this result shows that the cavity HHG setup can be used to engineer controlled multi-mode correlations.

The diagonal of this map corresponds to the single-mode second-order correlation function $g_2$ in Eq.\eqref{Eq:g2wans}. As can be seen, the majority of the modes possesses super-Poissonian statistics $(1<g_2<2)$, as expected due to the dominant semiclassical contribution to emission. However, some modes in the dip of the far-field spectrum experience destructive interference, which leads to sub-Poissonian statistics.

As a result, the simple 1D SSH model reveals an important outcome: quantum geometric properties of the material play a primary role in generating high-harmonic light, determining the quantum properties of the emitted light at the single-mode level and controlling multi-mode correlations.

\section{Conclusion}
\label{sec:conclusion}

To conclude, we have developed the theory connecting the quantum statistical properties of radiation nonlinearly scattered of the arbitrary crystal to the two-time current correlation functions for the classically driven crystal. We note that while in the Manuscript we only applied the formalism to the systems of non-interacting electrons, it can be straightforwardly applied to the systems of interacting electrons provided that two time current-current correlations are obtained using either diagrammatic techniques or with time-dependent density functional theory (TD-TDFT) methods. The developed theory paves the way towards novel methods of probing electron quantum correlations with quantum optical set-ups and towards possibility of quantum light engineering with correlated quantum phases of matter.

\section{Acknowledgments}
We are grateful to M. Ciappina for useful discussions. I.I. acknowledges the support of the Natural Sciences and Engineering Research Council of Canada (NSERC) [Discovery Grant No. 2024-05599].
D.I. and A.S. were supported by the Sydney Quantum Academy, Sydney, NSW, Australia. D.I. also acknowledges UTS for the support through the International Research Scholarship. 

\newpage
\appendix

\section{Derivation of density matrices from the universal Lindblad-like
equation}

\subsection{Matter subsystem}
\label{app:A1}

We start from the weak-correlation expansion for the reduced
density matrices of the matter and photonic subsystems, namely

\begin{equation}
    \begin{aligned}
        \dot{\rho}_{M}(t) = &-i\left[\mathrm{Tr}_{F}[H_{eff}(t)\rho_{F}(t)\right], \rho_{M}(t)] \\
-& \mathrm{Tr}_{F}\Bigl[H_{eff}(t), \int\limits_{-\infty}^{t} dt'[\tilde{H}_{eff}(t'), \rho_{M}(t') \otimes \rho_{F}(t')]\Bigr], \\
\dot{\rho}_{F}(t) = &-i\left[\mathrm{Tr}_{M}[H_{eff}(t)\rho_{M}(t)\right], \rho_{F}(t)] \\
-& \mathrm{Tr}_{M}\Bigl[H_{eff}(t), \int\limits_{-\infty}^{t} dt'[\tilde{H}_{eff}(t'), \rho_{M}(t') \otimes \rho_{F}(t')]\Bigr],
    \end{aligned}\label{Eq:A1}
\end{equation}
where
\begin{equation}
\begin{aligned}
    \tilde{H}_{eff}(t')&=\hat{H}_{eff}(t')\\&-\text{Tr}_{M}[\hat{H}_{eff}(t')\rho_M(t')]-\text{Tr}_{F}[\hat{H}_{eff}(t')\rho_F(t')].
\end{aligned}\label{Eq:A2}
\end{equation}

By one integration over the time of matter reduced density matrix we obtain an integral equation, which emphasizes the non-Markovian nature of the equations. 
\begin{equation}
    \begin{aligned}
        &\rho_{M}(t) = \rho_0-i\int\limits_{-\infty}^{t} dt'\left[\mathrm{Tr}_{F}[H_{eff}(t')\rho_{F}(t')\right], \rho_{M}(t')] \\&
- \mathrm{Tr}_{F} \Bigl[\int\limits_{-\infty}^{t} dt'H_{eff}(t'), \\ & \ \ \ \ \ \ \ \ \ \ \ \ \ \ \ \ \ \ \ \ \int\limits_{-\infty}^{t'} dt''[\tilde{H}_{eff}(t''), \rho_{M}(t'') \otimes \rho_{F}(t'')]\Bigr].
    \end{aligned}\label{Eq:A3}
\end{equation}
As mentioned in the main text, in this research we study the weak light-matter coupling regime when $N^2g_0\leq1$. This allows us to sufficiently simplify the equations under consideration and obtain a relatively simple analytical solution. Moreover, because we are interested only in photonic statistics, it further simplifies the reduced density matrix equation, since we need to preserve only the terms that contribute to the non-vanishing orders of light-matter interaction in the photonic equations. 

Now we consider each term separately. The first term corresponds to the unperturbed initial condition of the material system, which means that in the limit of vanishing light-matter coupling the interaction representation in which this equation is written becomes nothing but the Heisenberg representation. The second term denotes unitary dynamics, which arise from the finite light-matter interaction between a cavity and a material system. By definition, it is proportional to the average field $\langle A^{(\mu)}_{Q}(t)\rangle_F$. According to the reduced photonic density matrix equation~\eqref{Eq:A1}, it can be noticed that the average field is proportional to at least $g^2_0N$ because $\langle A^{(\mu)}_{Q}(t)\rangle_F\sim g^2_0\langle j_{\mu}(t)\rangle\sim g^2_0N$ at the leading nonvanishing order in the light-matter coupling. Such scaling allows us to omit the time-dependence of $\rho_M(t')$ in the second term in Eq.~\eqref{Eq:A3} and replace it with the  unperturbed value $\rho_0$.  The third term in Eq.\eqref{Eq:A3} is proportional to non-unitary Lindblad-like dynamics. By using the same evaluation strategy, one can show that it is proportional only to $g^2_0$ which in turn is out of scope for the considered photonic equation. Hence, within the approximation used for the photonic equations, the third term can be omitted entirely. This results in the already familiar Eq.~\eqref{Eq:rhoM} from the main text 
\begin{equation}
\begin{aligned}
    \rho_M(t)&=\rho_0-i\int\limits_{-\infty}^{t} dt'\left\langle A^{(\mu)}_Q(t')\right\rangle_F [j_{\mu}(t'),\rho_0].
    \end{aligned}\label{Eq:A4}
\end{equation}
To make our conclusion about the scaling of the third term more explicit, one can study the average paramagnetic current in a solid system by directly using Eq.\eqref{Eq:A3} with the time-dependent density matrix terms $\rho_M(t')$ omitted on the RHS. 
\begin{widetext}
\begin{equation}
\begin{aligned}
    \langle j_{\mu}(t)\rangle=\langle j_{\mu}(t)\rangle_0+\int\limits_{-\infty}^{t} dt' \chi_{\mu,\nu}(t,t')\left\langle A^{(\nu)}_Q(t')\right\rangle_F&-\frac{1}{2}\int\limits_{-\infty}^{t} dt'\int\limits_{-\infty}^{t'} dt'' \left\langle[[j_{\mu}(t),j_{\nu}(t')],j_{\nu}(t'')]\right\rangle_0
     \left\langle \{\delta A^{(\nu)}_Q(t'),\delta A^{(\nu)}_Q(t'')\}\right\rangle_F\\&-\frac{1}{2}\int\limits_{-\infty}^{t} dt'\int\limits_{-\infty}^{t'} dt'' \left\langle\{[j_{\mu}(t),j_{\nu}(t')],\delta j_{\nu}(t'')\}\right\rangle_0
     \left\langle [ A^{(\nu)}_Q(t'), A^{(\nu)}_Q(t'')]\right\rangle_F,
\end{aligned}\label{Eq:A5}
\end{equation}
\end{widetext}
where $\delta O=O-\langle O\rangle$, the average photonic field commutator is just a number and is proportional to $g^2_0$, whereas the current double commutator grows linearly with $N$ for translationally invariant systems because
\begin{equation}
\begin{aligned}
    &\left\langle[[j_{\mu}(t),j_{\nu}(t')],j_{\nu}(t'')]\right\rangle_0\\ &=\sum_{\textbf{k}}\left\langle[[j_{\mu,\textbf{k}}(t),j_{\nu,\textbf{k}}(t')],j_{\nu,\textbf{k}}(t'')]\right\rangle_0\sim N.
\end{aligned}\label{Eq:A6}
\end{equation}
See the similar detailed derivation of Eq.~\eqref{Eq:A8}. According to Eq.\eqref{Eq:A5}, $\delta j\sim g^2_0N^2$, whereas $\delta A\sim g_0^2N^2$, check the discussion of Eq.\eqref{Eq:B12}. As a result, the third and forth terms in Eq.~\eqref{Eq:A5} are proportional to $g^4_0N^5$ whose contribution can therefore be neglected to the photonic dynamics. Hence, we finally arrive at Eq.\eqref{Eq:current} from the main text. 

\subsection{Photonic subsystem}
\label{app:A2}

We can rewrite the photonic density matrix equation~\eqref{Eq:A1} in the form of Eq.~\eqref{Eq:generalULL} by expanding the commutators in the Lindbladian term. This equation can be simplified further. Taking into account our discussion of the average paramagnetic current from the previous subsection, we can say that the unitary term in Eq.~\eqref{Eq:generalULL} is 
\begin{equation}
\begin{aligned}
    \langle\hat{H}_{eff}(t)\rangle_M&=A^{(\mu)}_Q(t)\left[\langle j_{\mu}(t)\rangle_0+\frac{1}{2}\langle j_{d,\mu\nu}(t)\rangle_0A^{(\nu)}_Q(t)\right]\\&+A^{(\mu)}_Q(t)\int\limits_{-\infty}^{t}dt'\chi_{\mu,\nu}(t,t')\langle A^{(\nu)}_Q(t')\rangle,\end{aligned}\label{Eq:A7}
\end{equation}
where the first line originates from the unitary evolution of $\hat{H}_{eff}$ and the second line is a consequence of the weak correlation regime between photonic fields and a solid system. 

Regarding the Lindbladian term, as noted in the main text, to determine the susceptibility and matter correlation tensor one can use $\rho_0$ instead of $\rho_M(t)$  for the same reason as in the analysis of the reduced matter density matrix equation. Initially, the Lindbladian term contains the two-current correlation as $\langle j_{\mu}(t)j_{\nu}(t')\rangle_0$ which grows quadratically with the number of emitters $N$. However, one can notice that 
\begin{equation}
\begin{aligned}
    &\langle j_{\mu}(t)j_{\nu}(t')\rangle_M=\text{Tr}\left[\sum\limits_{k_1,k_2}j_{\mu,k_1}(t)j_{\nu,k_2}(t')\left(\otimes_{k}\rho_k(t')\right)\right]\\&=\sum\limits_{k_1\neq k_2}\langle j_{\mu,k_1}(t)\rangle\langle j_{\nu,k_2}(t')\rangle+\sum\limits_{k}\langle j_{\mu,k}(t) j_{\nu,k}(t')\rangle\\&=\langle j_{\mu}(t)\rangle\langle j_{\nu}(t')\rangle+D_{\mu,\nu}(t,t'),
\end{aligned}\label{Eq:A8}
\end{equation}
where we defined a correlation matter tensor as
in Eq.~\eqref{Eq:D_tensor_k} in the main text. As a result, it is clear that the correlation matter tensor grows linearly with $N$. In the following discussion of photonic statistics, the scaling of the Lindbladian terms is discussed in detail as well as the nature of additional dynamics that are governed by it.

\section{Floquet problem}
\label{app:FP}
In this section, we introduce the Floquet formalism, which can be used to solve the dynamics of a periodically modulated material system and as a result to obtain the correlation matter tensor. Here and throughout the following analysis, we restrict our attention
to translationally invariant systems whose density matrices factorize over electron momentum. The general formalism is considered in~\cite{PhysRevA.92.062108}. The influence of the thermal bath is considered in the next part. 

One can write the Hamiltonian of the periodically modulated system in the band basis as
\begin{equation}
    \hat{H}_{sys}(t)=\sum\limits_{n}e^{-in\Omega t}\sum\limits_{\textbf{k}}\sum\limits_{p,m}h^{(n)}_{pm}(\textbf{k})c^{\dagger}_{p,\textbf{k}}c_{m,\textbf{k}}.
\end{equation}\label{Eq:FP1}
According to the Floquet theorem~\cite{Kamenev2023}, solutions of the time-dependent Schrödinger equation can be found using the anzatz 
\begin{equation}
    \psi_m(\textbf{k},t)=e^{-i\epsilon_{\alpha,\textbf{k}}t}\phi^{\alpha}_m(\textbf{k},t),
\end{equation}\label{Eq:FP2}
where $\epsilon_{\alpha,\textbf{k}}$ is the Floquet quasienergy, $\phi(t)=\phi(t+T)$ is a periodic function. These functions form a complete orthonormal
basis and index $\alpha=1,..,N_{band}$ labels the bands, where $N_{band}$ is the number of bands. We choose the quasienergies in the Floquet zone $0\leq\epsilon_{\alpha,\textbf{k}}<\Omega$ in order to be well-defined. One can expand the wave function $\phi$ in a Fourier series as $\phi^{\alpha}_m(\textbf{k},t)=\sum_{n}\phi^{\alpha,(n)}_m(\textbf{k})e^{-in\Omega t}$. As a result, time-dependent problem can be rewritten as an eigenvalue problem 
\begin{equation}
    \sum\limits_{n'=-N_{\Omega}}^{N_{\Omega}-1}\sum\limits_{m}h^{(n-n')}_{pm}(\textbf{k})\phi^{\alpha,(n')}_{m}(\textbf{k})=(\epsilon_{\alpha,\textbf{k}}+n\Omega)\phi^{\alpha,(n)}_{p}(\textbf{k}),\label{Eq:FP3}
\end{equation}
where $N_{\Omega}$ is the frequency cutoff. As can be seen, this equation must hold for every momentum $\textbf{k}$ and hence it can be solved independently for each $\textbf{k}$. In what follows, we henceforth suppress the explicit momentum dependence in the
notation. As a result, the corresponding Keldysh action can be written as
\begin{equation}
\begin{aligned}
    S=\sum\limits_{\textbf{k}}\int dt\sum\limits_{pm}\Bigg[&\overline{\psi}_{1,p}\left(i\partial^{R}_t\delta_{pm}-H_{pm}(t)\right)\psi_{1,m}(t)\\+&\overline{\psi}_{2,p}\left(i\partial^{A}_t\delta_{pm}-H_{pm}(t)\right)\psi_{2,m}(t)\Bigg],
    \end{aligned}\label{Eq:FP4}
\end{equation}
where $\psi_a,m(t)$ are Grassmann fields, $a=1,2$ is Keldysh index, and superscript $R(A)$ specifies the retarded (advanced) regularization of the discrete time derivative. We transform the action to the Floquet basis by introducing
two new Grassmann fields as $\chi^{\alpha}_a(t)=\sum_m\overline{\phi}^{\alpha}_m(t)\psi_{a,m}(t)$ and $\overline{\chi}^{\alpha}_a(t)=\sum_m\overline{\psi}_{a,m}(t)\phi^{\alpha}_m(t)$. The action then simplifies to 
\begin{equation}    
\begin{aligned}
    S_{\textbf{k}} = \int dt\sum\limits_{\alpha}\bigg[&\overline{\chi}^{\alpha}_{1}\left(i\partial^{R}_t-\epsilon_{\alpha,\textbf{k}}\right)\chi^{\alpha}_{1}(t)\\+&\overline{\chi}^{\alpha}_{2}\left(i\partial^{A}_t-\epsilon_{\alpha,\textbf{k}}\right)\chi^{\alpha}_{2}(t)\bigg].
\end{aligned}\label{Eq:FP5}
\end{equation}
Note that, in the Floquet basis, the action is diagonal. Finally, the Green function in this basis is 
\begin{equation}
    \begin{aligned}
        G^{R}_{0,\alpha\beta}(t,t')&=-i\langle\chi^{\alpha}_1(t)\overline{\chi}^{\beta}_1(t')\rangle\\&=-i\theta(t-t')e^{-i\epsilon_{\alpha,\textbf{k}}(t-t')}\delta_{\alpha\beta}=G^R_{0,\alpha}(t-t').
    \end{aligned}\label{Eq:FP6}
\end{equation}
The Keldysh function can be expressed in terms of the distribution
function $F_{\alpha\beta}(t,t')$ as
\begin{equation}
    \begin{aligned}
        G^{K}_{\alpha\beta}&(t,t')=\int dt''\\\times&\bigg[G^{R}_{\alpha}(t-t'')F_{\alpha\beta}(t'',t')-F_{\alpha\beta}(t,t'')G^{R}_{\beta}(t''-t')\bigg].
    \end{aligned}\label{Eq:FP7}
\end{equation}
In the absence of a relaxation mechanism, an electronic system can
remain in a coherent superposition. This coherence is represented by nonzero off-diagonal elements of
the distribution matrix, namely
\begin{equation}
    \begin{aligned}
    F_{0,\alpha\beta}&=\delta_{\alpha\beta}-2\sum\limits_{p}f_{p,\textbf{k}}\overline{\phi}^{\alpha}_p(\textbf{k},t_0)\phi^{\beta}_p(\textbf{k},t_0)\equiv\delta_{\alpha\beta}-2f^{_0}_{\alpha\beta}, \\
    F_{0,\alpha\beta}&(t,t')=\delta(t-t')F_{0,\alpha\beta}e^{-i(\epsilon_{\alpha,\textbf{k}}-\epsilon_{\beta,\textbf{k}})t'},
    \end{aligned}\label{Eq:FP8}
\end{equation}
where $t_0$ is the time at which pumping is switched on. For simplicity, we set $t_0=0$. Hence, the Keldysh correlation function is 
\begin{equation}
    G^K_{0,\alpha\beta}(t,t')=-ie^{-i\epsilon_{\alpha,\textbf{k}}t}F_{0,\alpha\beta}e^{i\epsilon_{\beta,\textbf{k}}t'}.\label{Eq:FP9}
\end{equation}
As a result, we are able to calculate any observable we need to characterize the emission spectrum. The average current in the system can be written as 
\begin{equation}
    \begin{aligned}
        \langle j_{\mu}(t)\rangle&=\sum_{pm,\textbf{k}}j^{\mu}_{pm,\textbf{k}}(t)\langle c^{\dagger}_{p,\textbf{k}} (t)c_{m,\textbf{k}}(t)\rangle \\&=-i\sum_{pm,\textbf{k}}j^{\mu}_{pm,\textbf{k}}(t)G^{<}_{mp,\textbf{k}}(t,t),
    \end{aligned}\label{Eq:FP10}
\end{equation}
where the current operator is $\hat{j}_{\mu}=\partial(\hat{H}_{sys}(t)-\hat{H}_0)/\partial A^{\mu}(t)$. One can rotate the lesser (greater) function to the Floquet basis as
\begin{equation}
    \begin{aligned}
        G^{<(>)}_{mp}(t,t')&=\sum\limits_{\alpha\beta}\phi^{\alpha}_m(t)\overline{\phi}^{\beta}_p(t')G^{<(>)}_{\alpha\beta}(t,t').
    \end{aligned}\label{Eq:FP11}
\end{equation}
One can note that the lesser function can be defined through $G^{R/A/K}$ as $G^{<(>)}=\frac{1}{2}\left[G^{K}\mp(G^R-G^A)\right]$ and hence 
\begin{equation}
    \begin{aligned}
        \langle j_{\mu}(t)\rangle&=\sum_{\alpha\beta,\textbf{k}}f^{_0}_{\alpha\beta}J^{\mu}_{\beta\alpha}(t)\\ J^{\mu}_{\beta\alpha}(t)&=\sum_{pm}\overline{\psi}_{\beta,p}(t) j^{\mu}_{pm,\textbf{k}}(t)\psi_{\alpha,m}(t),
    \end{aligned}\label{Eq:FP12}
\end{equation}
where in the second line we introduce a current operator in the Floquet basis.

\begin{widetext}
Proceeding similarly, one can find the correlation matter tensor  $D_{\mu,\nu}(t,t')$ as
\begin{equation}
    D_{\mu,\nu}(t,t')=\sum\limits_{\textbf{k}}\sum\limits_{\substack{pm \\ p'm'}}j^{\mu}_{pm,\textbf{k}}(t)j^{\nu}_{p'm',\textbf{k}}(t')\left[\left\langle c^{\dagger}_{p,\textbf{k}}(t)c_{m,\textbf{k}}(t)c^{\dagger}_{ p',\textbf{k}}(t')c_{m',\textbf{k}}(t')\right\rangle-\left\langle c^{\dagger}_{p,\textbf{k}}(t)c_{ m,\textbf{k}}(t)\right\rangle\left\langle c^{\dagger}_{p',\textbf{k}}(t')c_{ m',\textbf{k}}(t')\right\rangle\right]\label{Eq:FP13}
\end{equation}
\end{widetext}
Because the averaging is Gaussian, Wick’s theorem can be used to
evaluate the expectation value of the product of four operators as
\begin{equation}
    \langle c^{\dagger}_pc_mc^{\dagger}_{p'}c_{m'}\rangle=\langle c^{\dagger}_pc_m\rangle\langle c^{\dagger}_{p'}c_{m'}\rangle+\langle c^{\dagger}_pc_{m'}\rangle\langle c_mc^{\dagger}_{p'}\rangle,\label{Eq:FP14}
\end{equation}
where the two sign changes arising from fermionic anti-commutation
have been taken into account. As can be seen, the first term is the disconnected term, which cancels exactly with the second term in Eq.\eqref{Eq:FP13}. As a result, we obtain 
\begin{equation}
\begin{aligned}
    D^{\textbf{k}}_{\mu,\nu}(t,t')=&\sum\limits_{\substack{pm \\ p'm'}}j^{\mu}_{pm,\textbf{k}}(t)j^{\nu}_{p'm',\textbf{k}}(t')\\\times&\left\langle c^{\dagger}_{p,\textbf{k}}(t)c_{ m',\textbf{k}}(t')\right\rangle\left\langle c_{m,\textbf{k}}(t)c^{\dagger}_{ p',\textbf{k}}(t')\right\rangle   
\end{aligned}\label{Eq:FP15}
\end{equation}
Then we change the basis to move to the Grassmann fields $\chi^{\alpha}_{a}$. Moreover, one can notice that these averages are, by definition, the lesser and greater Green
functions, respectively. Hence, each of them can be decomposed via $\phi$ fields through Eq.\eqref{Eq:FP11}.
As a result, after some algebra we obtain correlation matter tensor in the following form
\begin{equation}
    \begin{aligned}
        D^{\textbf{k}}_{\mu,\nu}(t,t')=\sum\limits_{\substack{\alpha\beta \\ \alpha'\beta'}}f^{_0}_{\alpha\beta}\left(\delta_{\alpha'\beta'}-f^{_0}_{\alpha'\beta'}\right)J^{\mu}_{\beta\alpha'}(t)J^{\nu}_{\beta'\alpha}(t').
    \end{aligned}\label{Eq:FP16}
\end{equation}
In this form, the correlation matter tensor can be calculated straightforwardly. 

The developed method is based on dynamics without relaxation, an approximation that does not
generally hold in realistic systems. To take into account relaxation, one needs to determine the evolution of the distribution matrix by solving the
corresponding kinetic equation
\begin{equation}
    \begin{aligned}
        \bigg[\partial_t+\partial_{t'}+i(\epsilon_{\alpha,\textbf{k}}-\epsilon_{\beta,\textbf{k}})\bigg]F_{\alpha\beta}(t,t')=I_{coll}[F].
    \end{aligned}\label{Eq:FP17}
\end{equation}
Note that in the collisionless case Eq.\eqref{Eq:FP8} satisfies the above equation.

In the next part, we consider a particular collision mechanism, which allows us to obtain a self-consistent theory of a periodically driven electron
system that is initially in thermal equilibrium.

\section{Floquet kinetic equation}
\label{app:FKE}

In this section, we are going to derive the kinetic equation for electrons in a translationally invariant solid-state system coupled to thermal photonic bath. The electron system is modulated by periodic Floquet pumping. A more general setup was considered in~\cite{PhysRevA.92.062108}. In the work, the authors take into account only electron-electron interaction. They show that the influence of relaxation processes associated with such interaction vanishes as the system reaches a stationary Floquet state. Here, we neglect these processes and focus on the influence of an external
thermal bath. 

We start by writing the standard model of the interaction between the solid-state system and the photonic bath, where the material system is modulated periodically in time
\begin{equation}
\begin{aligned}
\hat{H}(t)&=\hat{H}_{sys}(t)+\hat{H}_{bath}+\hat{H}_{int}, \\  \hat{H}_{bath}&=\sum\limits_{q,\mu}\omega_na^{\dagger}_{\mu,q}a_{\mu,q}, \\
\hat{H}_{int,\textbf{k}}&=\sum\limits_{q,\mu}\sum\limits_{ pm}g^{q,\mu}_{pm}(t)c^{\dagger}_{p,\textbf{k}}c_{m,\textbf{k}}\left(a_{\mu,q}+a^{\dagger}_{\mu,q}\right),
\end{aligned}\label{Eq:FKE1}
\end{equation}
where $g^{q,\mu}_{mp}=g_{q}j^{\mu}_{mp}(t)$ is the bath-matter coupling strength, $g_n$ is the common bath
form factor. The corresponding electron-bath interaction term in the action can be written as
\begin{equation}
    \begin{aligned}
        S_{int,\textbf{k}}=-\int dt &\sum_{q,\mu}\sum_{pm}\sum_{ab}\sum_{s}\\&g^{q,\mu}_{pm}(t)\overline{\psi}_{a,p}(t)\hat{\gamma}^s_{ab}\psi_{b,m}(t)\left(a^{s}_{\mu,q}+\overline{a}^{s}_{\mu,q}\right),
    \end{aligned}\label{Eq:FKE2}
\end{equation}
where $a,b=1,2$ are Keldysh indexes for the fermion fields, $s=cl,q$ is Keldysh indexes for the boson field. To use the notation of the previous section, we rewrite this action in terms of the fermion fields $\chi^{\alpha}_a(t)$ to obtain
\begin{equation}
    \begin{aligned}
        S_{int,\textbf{k}}=-\int dt &\sum_{q,\mu}\sum_{\alpha\beta}\sum_{ab}\sum_{s}\\&g_{q}V^{\mu}_{\alpha\beta}(t)\overline{\chi}^{\alpha}_{a}(t)\hat{\gamma}^s_{ab}\chi^{\beta}_{b}(t)\left(a^{s}_{\mu,q}+\overline{a}^{s}_{\mu,q}\right),
    \end{aligned}\label{Eq:FKE3}
\end{equation}
where 
\begin{equation}
    V^{\mu}_{\alpha\beta}(t)=\sum\limits_{pm}\overline{\phi}^{\alpha}_p(t)j^{\mu}_{pm}(t)\phi^{\beta}_{m}(t).\label{Eq:FKE4}
\end{equation}
Our goal is to calculate the bath-induced fermionic self-energy. However, one can see that the interaction part is bilinear in terms of the fermion fields and hence the integrating out the bosonic field produces an effective
electron-electron interaction that is fourth order in the fermionic
fields. Hence, the remaining fermionic integral cannot be evaluated as a Gaussian
integral without further approximation. Nevertheless, one can notice the Gaussian nature of the photonic bath and hence, because the bath is Gaussian, only the second cumulant contributes
to the cumulant expansion, provided that the first moment vanishes 
\begin{equation}
    \langle e^{iS}\rangle_B=e^{-\frac{1}{2}\left\langle S^2_B\right\rangle_B},
\end{equation}
which gives an interaction term proportional to $\chi^4$ as expected.
Then, one can use the non-crossing approximation (NCA). The use of this approximation is justified by taking into account the fact that the light-matter coupling is relatively weak and is controlled by the
fine-structure constant. This aspect makes the approximation analogous to the approximation justified
by Migdal’s theorem~\cite{AltlandSimons2010} that is typically applied to electron-phonon interaction. However, the latter relies on adiabaticity adiabaticity of the interaction rather than its smallness, which is not the case for photons. As a result, the action can be written in the following form
\begin{equation}
    \begin{aligned}
        S_{\textbf{k}}=\int dt &\int dt' \sum\limits_{\alpha,\alpha'}\overline{\chi}^{\alpha}(t)\hat{G}^{-1}_{\alpha\alpha'}(t,t')\chi^{\alpha'}(t'),\\
        \hat{G}^{-1}_{\alpha\alpha'}(t,t')&=\hat{G}^{-1}_{0,\alpha}(t-t')\delta_{\alpha\alpha'}-\hat{\Sigma}_{\alpha\alpha'}(t,t'),
    \end{aligned}\label{Eq:FKE5}
\end{equation}
where $\chi^{\alpha}(t)=(\chi^{\alpha}_1(t), \chi^{\alpha}_2(t))^T$ should be understood as a vector in Keldysh space and self-energy can be found as 
\begin{equation}
    \begin{aligned}
        \hat{\Sigma}_{\alpha\alpha'}(t,t')=&\sum\limits_{\mu,\beta\beta'}V^{\mu}_{\alpha\beta}(t)V^{\mu}_{\beta'\alpha'}(t')\\&\times\left(\hat{\gamma}^{s}\hat{G}_{\beta\beta'}(t,t')\hat{\gamma}^{s'}\right)\frac{i}{2}\hat{D}^{s's}_{0,\mu}(t'-t).
    \end{aligned}\label{Eq:FKE6}
\end{equation}
As can be seen, this self-energy corresponds to the self-consistent Born approximation (SCBA). In Eq.\eqref{Eq:FKE6}, the matrix $\hat{D}(\tau)$ is a bath propagator, and because of the boson nature its structure in frequency space is 
\begin{equation}
    \hat{D}_0(\omega)=\begin{pmatrix}
D^{K}_0(\omega) & D^{R}_0(\omega)\\
D^{A}_0(\omega) & 0
\end{pmatrix},\label{Eq:FKE7}
\end{equation}
where $D^R_0(\omega)=[D^A_0(\omega)]^*$, $J_+(\omega)=\eta\omega\Theta(\omega)$ and 
\begin{equation}
    \begin{aligned}    D^R_0(\omega)&=\sum\limits_{\mu,q}g^2_q\left[\frac{1}{\omega-\omega_n+i0}-\frac{1}{\omega+\omega_n+i0}\right]\\&\equiv\Delta^R(\omega)-i\pi\ \text{sgn}(\omega) J(|\omega|).        
    \end{aligned}\label{Eq:FKE8}
\end{equation}
Because the bath is in thermal equilibrium with itself we can write the Keldysh function by using the FDT as
\begin{equation}
    D^K_0(\omega)=-2\pi iJ(\omega)\left[1+2n_B(T,\omega)\right],\label{Eq:FKE9}
\end{equation}
where $n_B(T,\omega)$ is a=the boson occupation number. In this study, we assume an Ohmic bath, so for the relaxation rate one can write $J(\omega)=\eta\omega e^{-|\omega|/\omega_{UV}}$, where $\omega_{UV}$ is the upper frequency cutoff, fixed to $\omega_{UV}=200\Omega$. In this study, we fix $\eta=10^{-2}$. As can be immediately noticed, the bath relaxation rate $J_+$ depends nonlinearly on frequency, namely it vanishes at negative frequencies. In what follows, this property is crucial to avoid nonphysical results.

At the same time, the self-energy has the fermionic Keldysh structure, namely 
\begin{equation}
    \hat{\Sigma}_{\alpha\alpha'}(t,t')=\begin{pmatrix}
\Sigma^{R}_{\alpha\alpha'}(t,t') & \Sigma^{K}_{\alpha\alpha'}(t,t')\\
0 & \Sigma^{A}_{\alpha\alpha'}(t,t')
\end{pmatrix},\label{Eq:FKE10}
\end{equation}
where the retarded, advanced and Keldysh components are given
\begin{widetext}
\begin{equation}
    \begin{aligned}
        \Sigma^{R}_{\alpha\alpha'}(t,t')&=\frac{i}{2}\sum\limits_{\mu,\beta\beta'}V^{\mu}_{\alpha\beta}(t)V^{\mu}_{\beta'\alpha'}(t')\left[G^R_{\beta\beta'}(t,t')D^{K}_{0,\mu}(t'-t)+G^K_{\beta\beta'}(t,t')D^{A}_{0,\mu}(t'-t)\right]\equiv\Delta_{\alpha\alpha'}(t,t')-\frac{i}{2}\Gamma_{\alpha\alpha'}(t,t'); \\ 
        \Sigma^{A}_{\alpha\alpha'}(t,t')&=\frac{i}{2}\sum\limits_{\mu,\beta\beta'}V^{\mu}_{\alpha\beta}(t)V^{\mu}_{\beta'\alpha'}(t')\left[G^A_{\beta\beta'}(t,t')D^{K}_{0,\mu}(t'-t)+G^K_{\beta\beta'}(t,t')D^{R}_{0,\mu}(t'-t)\right]\equiv\Delta_{\alpha\alpha'}(t,t')+\frac{i}{2}\Gamma_{\alpha\alpha'}(t,t');\\
        \Sigma^{K}_{\alpha\alpha'}(t,t')&=\frac{i}{2}\sum\limits_{\mu,\beta\beta'}V^{\mu}_{\alpha\beta}(t)V^{\mu}_{\beta'\alpha'}(t')\bigg[G^K_{\beta\beta'}(t,t')D^{K}_{0,\mu}(t'-t)-\big(G^R_{\beta\beta'}(t,t')-G^A_{\beta\beta'}(t,t')\big)
        \left(D^{R}_{0,\mu}(t'-t)-D^{A}_{0,\mu}(t'-t)\right)\bigg],
    \end{aligned}\label{Eq:FKE11}
\end{equation}
\end{widetext}
where in the first two lines, we denote the Lamb-shift term by $\Delta_{\alpha\alpha'}$ and relaxation term by $\Gamma_{\alpha\alpha'}$. Before proceeding further with the calculation of the self-energy part, it is important to discuss the structure of the kinetic equation to determine how to
calculate the self-energy. By performing  a Wigner transformation~\cite{Kamenev2023}, we can define the transformed distribution function $F_{\alpha\alpha'}(t,\omega)$ for which the kinetic equation can be written as 
\begin{widetext}
\begin{equation}
    \begin{aligned}
        \partial_tF_{\alpha\alpha'}(t,\omega)=-i(\epsilon_{\alpha}-\epsilon_{\alpha'})F_{\alpha\alpha'}(t,\omega)+i\Sigma^K_{\alpha\alpha'}(t,\omega)-\frac{1}{2}\left\{\Gamma(t,\omega),F(t,\omega)\right\}_{\alpha\alpha'}-i\left[\Delta(t,\omega),F(t,\omega)\right]_{\alpha\alpha'},
    \end{aligned}\label{Eq:FKE12}
\end{equation}    
\end{widetext}
where $\{..,..\}$ and $[..,..]$ stand for anti-commutator and commutator operation respectively, $(t+t')/2\rightarrow t$ denotes the central time and $\omega$ is a Fourier frequency with respect to the relative time $\tau=t-t'$. In our work, we implement the on-shell approximation, namely we set $F_{\alpha\beta}(t,\omega)=F_{\alpha\beta}(t)\delta\left(\omega-\frac{\epsilon_{\alpha}+\epsilon_{\beta}}{2}\right)$ and then integrate over all frequencies to obtain
\begin{widetext}
\begin{equation}
    \begin{aligned}
        \partial_tF_{\alpha\alpha'}=-i(\epsilon_{\alpha}-\epsilon_{\alpha'})F_{\alpha\alpha'}+i\Sigma^K_{\alpha\alpha'}\left(t,\frac{\epsilon_{\alpha}+\epsilon_{\alpha'}}{2}\right)-\frac{1}{2}\sum\limits_{\beta}\left[\Gamma_{\alpha\beta}\left(t,\frac{\epsilon_{\beta}+\epsilon_{\alpha'}}{2}\right)F_{\beta\alpha'}+F_{\alpha\beta}\Gamma_{\beta\alpha'}\left(t,\frac{\epsilon_{\beta}+\epsilon_{\alpha}}{2}\right)\right],
    \end{aligned}\label{Eq:FKE13}
\end{equation}    
\end{widetext}
This approximation is valid because the characteristic relaxation rate is assumed to be much less than the smallest characteristic driving energy, namely $\Omega$, so we set $\Omega\gg\Gamma_M$. At the same time we can neglect the Lamb-shift term because of the weak nature of bath-matter coupling. Meanwhile, because it is included the in kinetic equation through commutator, t does not affect conservation of the total electron number. 

As can be seen in Eq.\eqref{Eq:FKE11}, all parts depend on the vertices in the same way. The vertex product in the Wigner representation is therefore
\begin{widetext}
\begin{equation}
\begin{aligned}
    V^{\mu}_{\alpha\beta}(t)V^{\mu}_{\beta'\alpha'}(t')&=\sum_{n_1,n_2}   V^{\mu}_{\alpha\beta}(n_1)V^{\mu}_{\beta'\alpha'}(n_2)e^{-i(n_1+n_2)\Omega t}e^{-i(n_1-n_2)\Omega\tau/2}
    \approx\sum_{n}   V^{\mu}_{\alpha\beta}(n)V^{\mu}_{\beta'\alpha'}(-n)e^{-in\Omega\tau},
\end{aligned}\label{Eq:FKE14}
\end{equation}    
\end{widetext}
where we neglect all fast oscillating terms by averaging over the central time, consistently with the slow
relaxation dynamics. 

Finally, we employ an approximation to replace the full electron Green function with the free one from Eq.\eqref{Eq:FP6}, which replaces the self-consistent Born approximation with the Born
approximation. This is essentially valid because the light-matter coupling is weak. Due to this simplification, after some algebra one can find 
\begin{widetext}
    \begin{equation}
        \begin{aligned}
            \Sigma^K_{\alpha\alpha'}(t)=-i\pi\eta\sum\limits_{n,\beta\beta'}V^{\mu}_{\alpha\beta}(n)V^{\mu}_{\beta'\alpha'}(-n)\left[\frac{\omega_LF_B(\omega_L)+\omega_RF_B(\omega_R)}{2}F_{\beta\beta'}(t)-\frac{\omega_L+\omega_R}{2}\delta_{\beta\beta'}\right],
        \end{aligned}\label{Eq:FKE15}
    \end{equation}
where $\omega_L=n\Omega+\epsilon_{\beta}-\epsilon_{\alpha}$, $\omega_R=n\Omega+\epsilon_{\beta'}-\epsilon_{\alpha'}$ and $F_B(\omega)$ is the boson distribution function. Moreover, we assume equal occupations of the two bath-photon polarizations. The relaxation term is given by
    \begin{equation}
        \begin{aligned}
            \Gamma_{\alpha\alpha'}(t)=i\left(\Sigma^R_{\alpha\alpha'}(t)-\Sigma^A_{\alpha\alpha'}(t)\right)=\pi\eta\sum\limits_{n,\beta\beta'}V^{\mu}_{\alpha\beta}(n)V^{\mu}_{\beta'\alpha'}(-n)\left[\frac{\omega_LF_B(\omega_L)+\omega_RF_B(\omega_R)}{2}\delta_{\beta\beta'}-\frac{\omega_L+\omega_R}{2}F_{\beta\beta'}(t)\right].
        \end{aligned}\label{Eq:FKE16}
    \end{equation}
\end{widetext}
As a result, the kinetic equation \eqref{Eq:FKE13} is fully determined via Eq.\eqref{Eq:FKE15} and Eq.\eqref{Eq:FKE16}. There is no convenient way to simplify it further, so we leave it as it is. 

Experiments typically use relatively short laser pulses. This is important because as is known~\cite{PhysRevLett.113.073901}, interband transitions are the main source of HHG in spectrum, but they are also highly sensitive to dephasing in the material. Hence, one should choose a pulse duration that is sufficiently short compared with the characteristic dephasing times in order to maximize the high-harmonic yield. As a result, we do not need to determine the new stationary state of the kinetic equation, because it does not generate the best high harmonic performance. Instead, we are interested in the solution after several cycles of the laser pulse, which requires solving the time-dependent differential equation. 

In the absence of pumping, the collision integral should vanish in thermal equilibrium. In this case, from Eq.\eqref{Eq:FP3} we find that indexes $\alpha,\beta$ label the crystal bands, whereas all Fourier components of the current are zero except the static one, i.e. $n=0$. Hence, we get that for any $\alpha, \beta$ the following equality holds
\begin{equation}
    \frac{1+n_B(\epsilon_{\beta}-\epsilon_{\alpha})}{n_B(\epsilon_{\beta}-\epsilon_{\alpha})}=\frac{f_{\alpha}}{1-f_{\alpha}}\frac{1-f_{\beta}}{f_{\beta}}.\label{Eq:FKE17}
\end{equation}
Due to the boson bath statistics this equation has a unique solution $f_{\alpha}=(1+e^{\beta_{th}(\epsilon_{\alpha}-\mu)})^{-1}$, which corresponds to the Fermi-Dirac statistics, as expected in thermal equilibrium. Meanwhile, we obtain that all off-diagonal terms vanish to satisfy this limit. This happens because in the thermal equilibrium the state is fully incoherent. Pumping drives the system out of thermal equilibrium by inducing
transitions involving the absorption and emission of integer numbers of photons. 

Finally, one can check that the obtained kinetic equation\eqref{Eq:FKE12} conserves the total number of electrons. To do this, one should notice that the RHS of the equation can be viewed as a sum of terms that are proportional to different powers of the distribution function, namely $I_{coll}\sim F^0+F^{1}+F^2$. Hence, to establish conservation, we must show that each term is zero independently. The linear component in powers of $F$ from Keldysh correlation part is equal in magnitude and opposite in sign to the corresponding
term from the anti-commutator, which can shown by a cyclic permutation of the band indexes. The remaining terms cancel because $\omega_x$ is antisymmetric under $n\leftrightarrow-n$ and permutation of the band indexes. 

Another important comment should be made about the LHS of the kinetic equation. In general, there is neither a static magnetic field nor a static electric field in the HHG problems. Hence, there are no terms that are proportional to Berry curvature or to electric field.

\section{Matter statistics}

As shown in Appendix~\ref{app:FP}, the average current and matter correlation tensor can be obtained from the lesser and greater Green functions. In this section, we are going to derive general expressions for observables that we need to study cavity field statistics. Due to the weak light-matter interaction strength we neglect vertex corrections in the evaluation of the current and current-current correlation function in this work.

\subsection{Current}
\label{app:D1}
We begin by calculating the average current. According to Eq.\eqref{Eq:FP10} and \eqref{Eq:FP12}, the average current is
\begin{equation}
\begin{aligned}
    \langle j_{\mu}(t)\rangle&=\sum\limits_{\alpha\beta,\textbf{k}}\left(-ie^{i\epsilon_{\alpha}t}G^{<}_{\alpha\beta}(t,t)e^{-i\epsilon_{\beta}t}\right)J^{\mu}_{\beta\alpha}(t)\\&=\sum\limits_{\alpha\beta,\textbf{k}}\left(e^{i\epsilon_{\alpha}t}f_{\alpha\beta}(t)e^{-i\epsilon_{\beta}t}\right)J^{\mu}_{\beta\alpha}(t),
\end{aligned}\label{Eq:cur1}    
\end{equation}
where $f_{\alpha\beta}(t)=(\delta_{\alpha\beta}-F_{\alpha\beta}(t))/2$. Because we are working within the on-shell approximation, one can write the kinetic equation directly to $f_{ab}$ as 
\begin{equation}
\begin{aligned}
    \partial_tf_{\alpha\beta}(t)&=-i(\epsilon_{\alpha}-\epsilon_{\beta})f_{\alpha\beta}(t)\\&-\frac{1}{2}\left\{\Gamma(t),f(t)\right\}_{\alpha\beta}+\Gamma^{in}_{\alpha\beta}(t),    
\end{aligned}
\end{equation}
where $\Gamma^{in(out)}_{\alpha\beta}=\frac{1}{2}\left[\Gamma\mp i\Sigma^K\right]_{\alpha\beta}$.
Hence, in the case of negligible coupling between the system and bath, only the rotating term survives and hence we get $f_{\alpha\beta}(t)=e^{-i(\epsilon_{\alpha}-\epsilon_{\beta})t}f^{_0}_{\alpha\beta}$ and the solution from Eq.\eqref{Eq:FP12} is recovered.

\subsection{Correlation matter tensor}
\label{app:D2}
One can immediately use our results from Eq.\eqref{Eq:FP15} to write the tensor in a form analogous to in Eq.\eqref{Eq:FP16} 
\begin{equation}
    \begin{aligned}
        D^{\textbf{k}}_{\mu,\nu}(t,t')=&\sum\limits_{\substack{\alpha\beta \alpha'\beta'}}\left(e^{i\epsilon_{\alpha}t'}G^{<}_{\alpha\beta}(t',t)e^{-i\epsilon_{\beta}t}\right)\\&\ \ \ \ \times \left(e^{i\epsilon_{\alpha'}t}G^{>}_{\alpha'\beta'}(t,t')e^{-i\epsilon_{\beta'}t'}\right)\\&\ \ \ \ \times J^{\mu}_{\beta\alpha'}(t)J^{\nu}_{\beta'\alpha}(t').
    \end{aligned}\label{Eq:TD1}
\end{equation}
Because we are working within the on-shell approximation, one can establish a relation between the lesser and greater functions in Wigner representation
\begin{equation}
    G^{>}_{\alpha\beta}(t,\omega)-G^{<}_{\alpha\beta}(t,\omega)\approx-2\pi i\delta_{\alpha\beta}\times\delta\left(\omega-\frac{\epsilon_{\alpha}+\epsilon_{\beta}}{2}\right).\label{Eq:TD3}
\end{equation}
Moreover, one can notice that the same argument can be used to simplify the lesser function as
\begin{equation}
\begin{aligned}
    G^{<}_{\alpha\beta}(t,\omega)&=2\pi if_{\alpha\beta}(t)\times\delta\left(\omega-\frac{\epsilon_{\alpha}+\epsilon_{\beta}}{2}\right).
\end{aligned}\label{Eq:TD4}
\end{equation}
As a result, the Wigner-transformed correlation matter tensor is 
    \begin{equation}
    \begin{aligned}
        D^{\textbf{k}}_{\mu,\nu}(t,\omega)&=2\pi\sum\limits_{\alpha\beta\alpha'\beta'}\sum\limits_{n_1,n_2}f_{\alpha\beta}(t)\left(\delta_{\alpha'\beta'}-f_{\alpha'\beta'}(t)\right)\\& \ \ \ \times V^{\mu}_{\beta\alpha'}(n_1)V^{\nu}_{\beta'\alpha}(n_2)e^{-i(n_1+n_2)\Omega t}\\& \ \ \  \times\delta\left(\omega-\frac{n_1-n_2}{2}\Omega-\frac{\epsilon_{\alpha'}+\epsilon_{\beta'}-\epsilon_{\alpha}-\epsilon_{\beta}}{2}\right). 
    \end{aligned}\label{Eq:TD5}     
\end{equation}
As can be noticed, the spectrum consists of sharp peaks because of the delta function. In principle, one can take into account the broadening due to the dissipation. However, this lies beyond the scope of this paper and may be studied in more detail in future work. Because the system-bath coupling is weak and the distribution matrix varies slowly, we can set correspondence of obtained tensor up to the first time derivative $\partial_tf(t)$ to it's time representation that is written as
\begin{equation}
\begin{aligned}
    D^{\textbf{k}}_{\mu,\nu}(t,t')&=\sum\limits_{\alpha\beta\alpha'\beta'}f_{\alpha\beta}\left(\overline{t}\right)\left(\delta_{\alpha'\beta'}-f_{\alpha'\beta'}\left(\overline{t}\right)\right)\\&  \times e^{i(\epsilon_{\alpha}-\epsilon_{\beta})\overline{t}}e^{i(\epsilon_{\alpha'}-\epsilon_{\beta'})\overline{t}}J^{\mu}_{\beta\alpha'}(t)J^{\nu}_{\beta'\alpha}(t'),
\end{aligned}\label{Eq:TD6} 
\end{equation}
where $\overline{t}=(t_1+t_2)/2$. One can immediately obtain the relaxationless limit from Eq.\eqref{Eq:FP16} by setting the occupation coefficients to their initial values, namely $f_{\alpha\beta}(t)\rightarrow f^{_0}_{\alpha\beta}$. 

As before, the case of zero pumping is of particular interest. In this situation, only $n=0$ contributions survive and $\alpha,\beta$ correspond to band numbers. In the absence of pumping, the system remains in its equilibrium state. After some algebra, we obtain
\begin{equation}
    \begin{aligned}
        D_{\mu,\nu}(\epsilon)&\approx\sum\limits_{pm,\textbf{k}}f_{p,\textbf{k}}(1-f_{m,\textbf{k}})\epsilon^2_{mp,{\textbf{k}}}r^{pm}_{\mu,\textbf{k}}r^{mp}_{\nu,\textbf{k}} \delta(\epsilon-\epsilon_{mp,\textbf{k}}),
    \end{aligned}\label{Eq:TD7}
\end{equation}
where we have restored the explicit momentum dependence and we introduce $r^{mp}_{\mu,{\textbf{k}}}=\langle u_{k,m}|i\partial_{\mu}u_{k,p}\rangle$ as the non-Abelian Berry connection in
momentum space, and $\epsilon_{mp}=\epsilon_m-\epsilon_p$.
If the temperature is zero, then there is no energy source in the system, which must result in zero cavity mode population. In the insulating phase considered here, to get nonzero $D$ we must assume at least $\epsilon_{mp,\textbf{k}}>0$, because at zero temperature $f_{p,\textbf{k}}=\theta(-\epsilon_{p,\textbf{k}})$ Hence, the spectral noise function is 
\begin{equation}
    S_{\mu,m}\sim D_{\mu\mu}(-\omega_m)\sim\delta(\omega_m+\epsilon_{mp,\textbf{k}}).\label{Eq:TD8}
\end{equation}
Because $\omega_m$ is always positive, the delta function condition is never satisfied, so $S_{\mu,m}(\omega_m)=0$. As a result, there is no generation of photons in the cavity due to the presence of the crystal inside it, which means that detailed balance is established.

\subsection{Relations between the correlation matter tensor and other well-known correlation functions}
\label{app:D3}

In this section, we establish relations between tensor $D_{\mu,\nu}$ with other correlation functions and macroscopic definitions. From the preceding analysis and a macroscopic perspective, we can immediately notice the following relation between the instantaneous susceptibility and the matter correlation tensor
\begin{equation}
    \chi_{\mu\nu}(\tau)=-i(D_{\mu\nu}(\tau)-D_{\nu\mu}(-\tau)),\label{Eq:TD18}
\end{equation}
which results in 
\begin{equation}
    \begin{aligned}
        \chi_{\mu,\nu}(\epsilon)&=-i(D_{\mu\nu}(\epsilon)-D_{\nu\mu}(-\epsilon))\\&=i\sum\limits_{\textbf{k}}\sum\limits_{pm}f_{pm,\textbf{k}}\epsilon^2_{pm,{\textbf{k}}}r^{mp}_{\mu,\textbf{k}}r^{pm}_{\nu,\textbf{k}}\delta(\epsilon-\epsilon_{pm,\textbf{k}}).
    \end{aligned}
\end{equation}
This is a well-known result of Kubo linear-response theory. 

From the macroscopic point of view, the average current and polarization obey the relation $\langle j_{\mu}(t)\rangle=\partial_t\langle d_{\mu}(t)\rangle$. Therefore, formally we can define the relation between current-current and dipole-dipole correlation functions as 
\begin{equation}
    \langle\hat{j}_{\mu}(t)\hat{j}_{\nu}(t')\rangle=\frac{\partial}{\partial t}\frac{\partial}{\partial t'}\langle\hat{d}_{\mu}(t)\hat{d}_{\nu}(t')\rangle.\label{Eq:AD7}
\end{equation}
For the instantaneous contribution, we have 
\begin{equation}
\begin{aligned}
        &\frac{\partial}{\partial t}\frac{\partial}{\partial t'}\langle\hat{d}_{\mu}(t)\hat{d}_{\nu}(t')\rangle=\frac{\partial}{\partial t}\frac{\partial}{\partial t'}\langle\hat{d}_{\mu}(t-t')\hat{d}_{\nu}(0)\rangle\\&=-    \frac{\partial^2}{\partial \tau^2}\langle\hat{d}_{\mu}(\tau)\hat{d}_{\nu}(0)\rangle.\label{Eq:AD8}
\end{aligned}
\end{equation}
As noted in~\cite{Verma}, the relevant dipole-dipole correlation function is
\begin{equation}
    \mathcal{Q}_{\mu\nu}(\tau)=\text{Tr}\left[\hat{P}\hat{r}_{\mu}(\tau)(1-\hat{P})\hat{r}_{\nu}(0)\right],\label{Eq:AD9}
\end{equation}
where $\hat{P}$ is the projector onto the ground state. Therefore, we deduce that in the stationary situation instantaneous correlation matter tensor is closely related to the quantum geometric tensor (QGT) as 
\begin{equation}
    D^{(0)}_{\mu,\nu}(\tau)=-\frac{\partial^2}{\partial\tau^2}\mathcal{Q}_{\mu,\nu}(\tau).\label{Eq:AD10}
\end{equation}
This relation manifests the influence of geometry and topology on photonic signatures. Moreover, Eq.~\eqref{Eq:AD7} provides a more general correspondence between the quantum geometric tensor that is nonlocal in time and
the correlation matter tensor and paves the way to define physically meaningful dipole-dipole correlations in translational invariant systems under periodic driving.

\section{Cavity field equations}
\label{app:B}

In this section, we  are going to derive in detail the photonic equations of motion when the solid is placed inside a one-sided cavity. We start from the boundary condition relating each of the far-field amplitudes outside the cavity to the intracavity field, so the outgoing field is
\begin{equation}
    a^{\mu}_{out}(t)=\sum\limits_{m}\sqrt{2\gamma_m}a_{\mu,m}(t)-a^{\mu}_{in}(t)\label{Eq:B1}
\end{equation}
As a result, interference terms between the input and the cavity field may contribute to the observed moments outside the cavity. Initially, in the rotating frame, there is no incoming field, so $\langle a^{\mu}_{in}(t)\rangle=0$, because the laser field is taken into account as a classical driving force for a solid system. 

First of all, we are going to derive the corresponding equation of motion for a single operator, namely the field-amplitude operator $a_{\mu,m}$ of the $m$-th mode. Because Eq.~\eqref{Eq:ull} is written in the interaction picture, we note that the total time derivative of the expectation value of any operator $O_H(t)$ in the Heisenberg picture is 
\begin{equation}
\begin{aligned}
    \frac{d\langle O_H(t)\rangle}{dt}&=\text{Tr}\left[ O_I(t)\frac{d\rho_F(t)}{dt}\right]+\text{Tr}\left[\frac{\partial O_I(t)}{\partial t}\rho_F(t)\right]\\&+\text{Tr}\left[\mathcal{L}_f[\rho_F(t)]O_I(t)\right],\label{Eq:B2}
\end{aligned}
\end{equation}
where the first line corresponds to the standard transformation from the interaction picture to the Heisenberg picture, while the second line describes leakage from the cavity into open space. We extract this term explicitly from $d\rho_F/dt$ to emphasize the open cavity considered here. Cavity leakage is described in the standard Lindbladian form as 
\begin{equation}
\mathcal{L}_f[\rho]=\sum_{\mu,m}\gamma_m\left(2a_{\mu,m}\rho a^{\dagger}_{\mu,m}-\rho a^{\dagger}_{\mu,m}a_{\mu,m}-a^{\dagger}_{\mu,m}a_{\mu,m}\rho\right)\label{Eq:B3}.
\end{equation}
Keeping this in mind and taking into account Eq.~\eqref{Eq:ull} and Eq.~\eqref{Eq:A7}, the unitary contribution to the evolution of the field amplitude can be written as 
\begin{equation}
    \begin{aligned}
        -i&\text{Tr}\left(a_{\mu,m}e^{-i\omega_m t}\left[\langle\hat{H}_{eff}\rangle_M,\rho_F(t)\right]\right)\\=&-i\frac{g_0}{\sqrt{\omega_m}}\langle j_{\mu}(t)\rangle_0 -i\frac{g_0}{\sqrt{\omega_m}}\int\limits_{-\infty}^{t} dt'\langle A^{(\nu)}_Q(t')\rangle \\&\times\bigg(\chi_{\mu,\nu}(t,t')+2\langle j_{d,\mu\nu}(t')\rangle_0\delta(t-t')\bigg).
    \end{aligned}\label{Eq:B4}
\end{equation}
The expression under the last integral is nothing but a matter conductivity tensor $\sigma_{\mu,\nu}(t,t')=-\Theta(t-t')\left(\langle j_{d,\mu\nu}(t')\rangle_0+\int_{t'}^{t}dt''\chi_{\mu,\nu}(t,t'')\right)$. Hence, the unitary evolution part can be rewritten as 
\begin{equation}
    \eqref{Eq:B4}=-i\frac{g_0}{\sqrt{\omega_m}}\left[\langle j_{\mu}(t)\rangle_0 +\int dt'\sigma_{\mu,\nu}(t,t')\langle E^{(\nu)}_Q(t')\rangle\right]\label{Eq:B5}, 
\end{equation}
where we define $E^{(\mu)}_{Q}(t)=-\partial  A^{(\mu)}_{Q}(t)/\partial t$ by analogy with the electric field operator. The physical meaning of each term is straightforward. The first one is a semiclassical response of the solid system under the classical pumping force, while the second term is the backaction of the emitted field on the solid which occurs because the emitted fields in the past induce extra current in the crystal through nonlocal conduction in time. Each of the terms has a simple physical meaning by analogy with classical electrodynamics. This is not surprising, especially since they govern the unitary evolution of the field.

Turning to the evolution governed by the Lindbladian term in Eq.\eqref{Eq:ull}, it is instructive to use the original form of the Lindbladian as

\begin{widetext}
\begin{equation}
    \begin{aligned}
        \text{Tr}&\left(a_{\mu,m}e^{-i\omega_m t}\left[j_{\mu}(t)A^{(\mu)}_Q(t),\left[j_{\nu}(t')A^{(\nu)}_Q(t')-\langle j_{\nu}(t')\rangle A^{(\nu)}_Q(t')-j_{\nu}(t')\langle A^{(\nu)}_Q(t')\rangle,\rho_0\otimes\rho_F(t')\right]\right]\right)\\ =&\text{Tr}\left(\left[\left[a_{\mu,m}e^{-i\omega_m t},j_{\mu}(t)A^{(\mu)}_Q(t)\right],j_{\nu}(t')A^{(\nu)}_Q(t')-\langle j_{\nu}(t')\rangle A^{(\nu)}_Q(t')-j_{\nu}(t')\langle A^{(\nu)}_Q(t')\rangle\right],\rho_0\otimes\rho_F(t')\right) \\ =&\frac{g_0}{\sqrt{\omega_m}}\text{Tr}\left(\left[j_{\mu}(t),\ j_{\nu}(t')A^{(\nu)}_Q(t')-\langle j_{\nu}(t')\rangle A^{(\nu)}_Q(t')-j_{\nu}(t')\langle A^{(\nu)}_Q(t')\rangle\right],\rho_0\otimes\rho_F(t')\right)\\=&\frac{g_0}{\sqrt{\omega_m}}\left(\langle [j_{\mu}(t),j_{\nu}(t')]\rangle_0\langle  A^{(\nu)}_Q(t')\rangle-0-\langle [j_{\mu}(t),j_{\nu}(t')]\rangle_0\langle  A^{(\nu)}_Q(t')\rangle\text{Tr}(\rho_F(t))\right)=0
    \end{aligned}\label{Eq:B6}
\end{equation}
\end{widetext}
This result follows from preservation of the trace of $\rho_F(t)$ over time. Thus, the average field amplitudes of the field in the weak correlation regime are not affected by the non-unitary evolution at all. This vanishing contribution supports replacing the matter density matrix by its unperturbed value in this term. The two remaining terms follow directly. The leakage term generates the $-\gamma_m\langle a_{\mu,m}(t)\rangle$ part, and the one with the partial derivative generates the $-i\omega_m\langle a_{\mu,m}(t)\rangle$ part. Overall, the equation of motion of the average amplitude inside the cavity is 
\begin{widetext}
\begin{equation}
    \begin{aligned}
        \frac{d\langle a_{\mu,m}(t)\rangle}{dt}&=-(\gamma_m+i\omega_m) \langle a_{\mu,m}(t)\rangle-i\frac{g_0}{\sqrt{\omega_m}}\langle j_{\mu}(t)\rangle_0 -i\frac{g_0}{\sqrt{\omega_m}}\int dt'\sigma_{\mu,\nu}(t,t')\langle E^{(\nu)}_Q(t')\rangle
    \end{aligned}\label{Eq:B7}.
\end{equation}
\end{widetext}
This agrees with Eq.\eqref{Eq:amplitude} of the main text. As stated in the main text, it is convenient to solve this equation by using the following ansatz 
\begin{equation}
    \langle a_{\mu,m}(t)\rangle=a^{(cl)}_{\mu,m}(t)+a^{(q)}_{\mu,m}(t)\sim g_0N(1+g^2_0N).\label{Eq:B8}
\end{equation}
The first term describes the semiclassical evolution, while the second describes backaction. This will be extremely helpful in understanding the origin of quantum properties of the emitted light. For the semiclassical response, the analytical solution is
\begin{equation}
    a^{(cl)}_{\mu,m}(\omega)=\frac{g_0}{\sqrt{\omega_m}}G^m_{c}(\omega)\langle j_{\mu}(\omega)\rangle_0,\label{Eq:B9}
\end{equation}
where $G^m_{c}(\omega)=(\omega-\omega_m+i\gamma_m)^{-1}$ is a cavity propagator. In principle, it is possible to rewrite Eq.\eqref{Eq:B8} in the form of Eq.\eqref{Eq:B9} by dressing the cavity propagator. In this case, the cavity self-energy is proportional to the conductivity tensor. However, this description lies beyond the scope of this paper.  

To determine photonic statistics we also need to evaluate observables involving two field operator, such as the average number of photons $n_{\mu,m}=\langle a^{\dagger}_{\mu,m}a_{\mu,m}\rangle$ and the average squared field $\langle a^2_{\mu,m}\rangle$ of the mode $m$. Deviations of these second moments from the corresponding products of mean  mplitudes indicate departures from
multi-mode coherent emission. We start from the mean mode populations. In the interaction picture the corresponding population operator has no explicit time dependence, so the contribution from the partial derivative is zero in the equation of motion. The leakage part produces the $-2\gamma_m n_{\mu,m}(t)$ term. The unitary evolution part provides analogous terms to those obtained in the field amplitude equation in Eq.\eqref{Eq:B5}, namely

\begin{widetext}
\begin{equation}
    \begin{aligned}
        -i\text{Tr}\left(a^{\dagger}_{\mu,m}a_{\mu,m}\left[\langle\hat{H}_{eff}\rangle_M,\rho_F(t)\right]\right)=\frac{1}{\omega_m}\langle E^{(\mu)}_Q(t)\rangle\left[\langle j_{\mu}(t)\rangle_0 +\int dt'\sigma_{\mu,\nu}(t,t')\langle E^{(\nu)}_Q(t')\rangle\right]\sim g^2_0N^2(1+g^2_0N).
    \end{aligned}\label{Eq:B10}
\end{equation}
Because it is governed only by the average amplitudes of the field, this term can generate only mixtures of coherent states up to the leading order of $g_0$. The most interesting part arises from the contribution of the non-unitary term in Eq.\eqref{Eq:ull}. We can calculate it as 

\begin{equation}
    \begin{aligned}
        -\text{Tr}&\left(a^{\dagger}_{\mu,m}a_{\mu,m}\left[j_{\mu}(t)A^{(\mu)}_Q(t),\left[j_{\nu}(t')A^{(\nu)}_Q(t')-\langle j_{\nu}(t')\rangle A^{(\nu)}_Q(t')-j_{\nu}(t')\langle A^{(\nu)}_Q(t')\rangle,\rho_0\otimes\rho_F(t')\right]\right]\right) \\ =&\frac{-i}{\omega_m}\text{Tr}\left(\left[j_{\mu}(t)E^{(\mu)}_{Q,m}(t),\ j_{\nu}(t')A^{(\nu)}_Q(t')-\langle j_{\nu}(t')\rangle A^{(\nu)}_Q(t')-j_{\nu}(t')\langle A^{(\nu)}_Q(t')\rangle\right],\rho_0\otimes\rho_F(t')\right)\\=&\frac{-i}{\omega_m}\left(\langle j_{\mu}(t)j_{\nu}(t')\rangle_0\langle E^{(\mu)}_{Q,m}(t)A^{(\nu)}_Q(t')\rangle-\langle j_{\nu}(t')j_{\mu}(t)\rangle_0\langle A^{(\nu)}_Q(t')E^{(\mu)}_{Q,m}(t)\rangle\right)\\+&\frac{i}{\omega_m}\left(\langle [j_{\nu}(t'),j_{\mu}(t)]\rangle_0\langle A^{(\nu)}_Q(t')\rangle\langle E^{(\mu)}_{Q,m}(t)\rangle+\langle j_{\nu}(t')\rangle\langle  j_{\mu}(t)\rangle_0\langle [E^{(\mu)}_{Q,m}(t),A^{(\nu)}_Q(t')]\rangle\right)
    \end{aligned}\label{Eq:B11}
\end{equation}
Any product of two operators can be decomposed into symmetric and antisymmetric parts as $\langle E^{(\mu)}_mA^{(\nu)}\rangle=\frac{1}{2}(\langle\{E^{(\mu)}_m,A^{(\nu)}\}\rangle+\langle[E^{(\mu)}_m,A^{(\nu)}]\rangle)$. Then we note that the commutator of two operators in this case is just a number, namely $\langle[E^{(\mu)}_m(t),A^{(\nu)}(t')]\rangle=ig^2_0\delta_{\mu,\nu}(e^{i\omega_m\tau}+e^{-i\omega_m\tau})e^{-\gamma_m|\tau|}$, where $\tau=t-t'$. Thus, it can be rewritten as

\begin{equation}
    \begin{aligned}
        \eqref{Eq:B11}&=\frac{g^2_0}{\omega_m}\left(D_{\mu,\mu}(t,t')e^{-i\omega_m\tau}+D_{\mu,\mu}(t',t)e^{i\omega_m\tau}\right)e^{-\gamma_m|\tau|}\\+&\frac{ig^2_0}{\omega_m}\chi_{\mu,\nu}(t,t')e^{-\gamma_m|\tau|}\left(\bigg[\left(\langle a^2_{\mu,m}(t')\rangle-\langle a_{\mu,m}(t')\rangle^2\right)+\left(\langle a^{\dagger}_{\mu,m}(t')a_{\mu,m}(t')\rangle-\langle a^{\dagger}_{\mu,m}(t')\rangle\langle a_{\mu,m}(t')\rangle\right)\bigg]e^{-i\omega_m\tau}-c.c\right).
    \end{aligned}\label{Eq:B12}
\end{equation}
\end{widetext}
The first line scales as $g^2_0N$, while the second part consists of differences between second- and first-order statistics. Its scaling will be discussed below. For now, we just omit this part. Returning to the first line, this term corresponds to the noise spectral function and as a result generates deviations from coherent radiation. Combining the terms, we can write the equation for the average population of the m-th mode as
\begin{widetext}
\begin{equation}
    \begin{aligned}
        \frac{dn_{\mu,m}(t)}{dt}&=-2\gamma_mn_{\mu,m}(t)+ \frac{d}{dt}\left[\langle  a^{\dagger}_{\mu,m}(t)\rangle\langle a_{\mu,m}(t)\rangle\right]+2\gamma_m\langle  a^{\dagger}_{\mu,m}(t)\rangle\langle a_{\mu,m}(t)\rangle+\frac{g^2_0}{\omega_m}\int\limits_{-\infty}^{\infty}d\tau D_{\mu,\mu}(t+\tau,t)e^{-i\omega_m\tau}e^{-\gamma_m|\tau|}\\&\equiv  -2\gamma_m\left(n_{\mu,m}(t)-n^{(coh)}_{\mu,m}(t)\right)+\frac{d}{dt}n^{(coh)}_{\mu,m}(t)+S_{\mu,m}(t). \end{aligned}\label{Eq:B13}
\end{equation}
\end{widetext}
One should note that when the temperature $T_{cav}$ of the optical bath is positive, the corresponding population of thermal bosons $\overline{n}_{th}$ should be added to the equation. However, in this work, we assume a sufficiently low temperature that the thermal population is negligible. The formal solution can be written as
\begin{equation}
    n_{\mu,m}(t)=n^{(coh)}_{\mu,m}(t)+\int\limits_{-\infty}^{t}dt'S_{\mu,m}(t')e^{-2\gamma_m(t-t')}.\label{Eq:B14}
\end{equation}
Several comments should be made here. Firstly, from the definition of the noise spectral function, it is clear that it grows as $S\sim g^2_0N$. Hence, the difference between the total population of each mode and its coherent contribution cannot grow faster than $g^2_0N$. Now we can return to Eq.~\eqref{Eq:B12} to estimate the scaling of the second line. It is clear that it is bounded by $g^2_0N\times g^2_0N\sim g^4_0N^2$ and hence it lies beyond the order retained here. 

By proceeding in the same way, the corresponding equation for the average squared field amplitude can be found. By introducing the new notation $\langle A,B\rangle=\langle AB\rangle - \langle A\rangle\langle B\rangle$, we can write the equation compactly as
\begin{widetext}
\begin{equation}
    \begin{aligned}
        \frac{d}{dt}\langle a_{\mu,m}(t),a_{\mu,m}(t)\rangle&=-2(\gamma_m+i\omega_m)\langle a_{\mu,m}(t),a_{\mu,m}(t)\rangle-\frac{g^2_0}{\omega_m}\left[i\langle j_{d,\mu\mu}\rangle+2\int\limits_{0}^{\infty} d\tau D_{\mu,\mu}(t+\tau,t) e^{-i\omega_m\tau}e^{-\gamma_m\tau}\right].    \end{aligned}\label{Eq:B15}
\end{equation}
\end{widetext}
As can be seen, the mean squared field differs from the square of the mean field by an additional noise contribution. This extra term is responsible for departures from coherence of the emitted light and for generating quantum properties of light. The formal solution for Eq.~\eqref{Eq:B15} can be found by analogy with Eq.~\eqref{Eq:B14}. 

The same procedure can be extended to more complicated observables that mix different modes, polarizations, etc. Turning to two-mode intracavity correlations, one needs to calculate the two-operator averages between different modes. To do so, we introduce the generalized noise correlation tensor $\Xi^{p_1p_2}_{\mu\nu}(t)$, defined by
\begin{equation}
\begin{aligned}
\Xi^{p_1p_2}_{\mu\nu}&(t)=\frac{2g^2_0}{\sqrt{\omega_{p_1}\omega_{p_2}}}\\\times&\Bigg(\frac{i\langle j_{d,\mu\nu}(t)\rangle}{2}+\int\limits_{0}^{\infty} d\tau D_{\mu,\nu}(t+\tau,t) e^{-i\omega_{p_2}\tau}e^{-\gamma_{p_2}\tau}\Bigg),\label{Eq:B16}
\end{aligned}
\end{equation}
Therefore, after some algebra, we can determine them as 
\begin{widetext}
\begin{equation}
\begin{aligned}
    \frac{d}{dt}\langle a_{\mu,p_1}(t),a_{\nu,p_2}(t)\rangle&=-(\gamma_{p_1}+\gamma_{p_2}+i(\omega_{p_1}+\omega_{p_2}))\ \langle a_{\mu,p_1}(t),a_{\nu,p_2}(t)\rangle-\frac{1}{2}\left[\Xi^{p_1p_2}_{\mu,\nu}(t)+\Xi^{p_2p_1}_{\nu,\mu}(t)\right],\\
    \frac{d}{dt}\langle a^{\dagger}_{\mu,p_1}(t),a_{\nu,p_2}(t)\rangle&=-(\gamma_{p_1}+\gamma_{p_2}+i(\omega_{p_2}-\omega_{p_1}))\ \langle a^{\dagger}_{\mu,p_1}(t),a_{\nu,p_2}(t)\rangle+\frac{1}{2}\left[\Xi^{p_1p_2}_{\mu,\nu}(t)+\Xi^{p_2p_1^*}_{\nu,\mu}(t)\right].
\end{aligned}\label{Eq:B17}
\end{equation}
\end{widetext}
By calculating the generalized noise correlation function between different modes, it is possible to determine the covariance matrix of the modes and hence estimate two-mode correlations.

\section{Photon field statistics}
\subsection{Intracavity statistics}
\label{app:C}
Despite the fact that far-field photon statistics are much easier to measure in practice than intracavity statistics, we first study the intracavity statistics because, as we know from Eq.\eqref{Eq:B1}, all outgoing fields are functions of the intracavity operators. As a result, we are interested only in particular forms of the intracavity statistics that are needed for the subsequent analysis of far-field statistics.

In this section, we study two correlation functions, namely the normal correlation $\langle a^{\dagger}_{\mu,p_1}(t)a_{\mu,p_2}(t+\tau)\rangle$ and the anomalous correlation $\langle a_{\mu,p_2}(t)a_{\mu,p_2}(t+\tau)\rangle$.The remaining correlations can be obtained from these two. Each can be calculated using the quantum regression theorem (QRT)~\cite{Zoller:1997su, Swain_1981,PhysRevA.99.033816, PhysRevA.109.052207}, which states that the two-time correlation function can be found by solving the same closed system of linear differential equations with modified initial conditions determined by the single-time observables. As a result, one can find solutions for both correlations as the sum of a homogeneous part that depends on the initial conditions and an inhomogeneous part.
\begin{widetext}
    \begin{equation}
        \begin{aligned}
            \langle a^{\dagger}_{\mu,{p_1}}(t)a_{\mu,p_2}(t+\tau)\rangle=&\langle a^{\dagger}_{\mu,p_1}(t)a_{\mu,p_2}(t)\rangle e^{-(\gamma_{p_2}+i\omega_{p_2})\tau}\\&+\frac{g^2_0}{\sqrt{\omega_{p_1}\omega_{p_2}}}\int\limits_{0}^{\tau}d\tau' e^{-(\gamma_{p_2}+i\omega_{p_2})(\tau-\tau')}\int\limits_{0}^{\infty}dse^{-(\gamma_{p_1}-i\omega_{p_1})s}\langle j_{\mu}(t-s)j_{\mu}(t+\tau')\rangle; \\
            \langle a_{\mu,{p_1}}(t)a_{\mu,p_2}(t+\tau)\rangle=&\langle a_{\mu,p_1}(t)a_{\mu,p_2}(t)\rangle e^{-(\gamma_{p_2}+i\omega_{p_2})\tau}\\&-\frac{g^2_0}{\sqrt{\omega_{p_1}\omega_{p_2}}}\int\limits_{0}^{\tau}d\tau' e^{-(\gamma_{p_2}+i\omega_{p_2})(\tau-\tau')}\int\limits_{0}^{\infty}dse^{-(\gamma_{p_1}+i\omega_{p_1})s}\langle j_{\mu}(t-s)j_{\mu}(t+\tau')\rangle.
        \end{aligned} \label{Eq:F1}
    \end{equation}
In this notation, the first term corresponds to the homogeneous solution, whereas the second pertains to the inhomogeneous part. It is convenient to rewrite everything in the frequency domain, so
\begin{equation}
    \begin{aligned}
        \text{(inhom)}_{norm}&=\frac{g^2_0}{\sqrt{\omega_{p_1}\omega_{p_2}}}\int\frac{d\omega_1}{2\pi}\int\frac{d\omega_2}{2\pi}e^{-i(\omega_1+\omega_2)t}\langle j(\omega_1)j(\omega_2)\rangle G^{p_2}_c(\omega_2)G^{p^*_1}_c(-\omega_1)\left[e^{-i\omega_2\tau}-e^{-(\gamma_{p_2}+i\omega_{p_2})\tau}\right]; \\
        \text{(inhom)}_{anom}&=\frac{g^2_0}{\sqrt{\omega_{p_1}\omega_{p_2}}}\int\frac{d\omega_1}{2\pi}\int\frac{d\omega_2}{2\pi}e^{-i(\omega_1+\omega_2)t}\langle j(\omega_1)j(\omega_2)\rangle G^{p_2}_c(\omega_2)G^{p_1}_c(\omega_1)\left[e^{-i\omega_2\tau}-e^{-(\gamma_{p_2}+i\omega_{p_2})\tau}\right].
    \end{aligned}\label{Eq:F2}
\end{equation}
From Eq.\eqref{Eq:B17} we can obtain the equal-time two-operator correlation functions in the frequency domain as 
\begin{equation}
    \begin{aligned}
        \langle a^{\dagger}_{\mu,p_1}(t)a_{\mu,p_2}(t)\rangle&=\frac{g^2_0}{\sqrt{\omega_{p_1}\omega_{p_2}}}\int\frac{d\omega_1}{2\pi}\int\frac{d\omega_2}{2\pi}e^{-i(\omega_1+\omega_2)t}\langle j(\omega_1)j(\omega_2)\rangle G^{p_2}_c(\omega_2)G^{p^*_1}_c(-\omega_1);\\
        \langle a_{\mu,p_1}(t)a_{\mu,p_2}(t)\rangle&=\frac{g^2_0}{\sqrt{\omega_{p_1}\omega_{p_2}}}\int\frac{d\omega_1}{2\pi}\int\frac{d\omega_2}{2\pi}e^{-i(\omega_1+\omega_2)t}\big(\langle j(\omega_1)j(\omega_2)\rangle +i\chi_{\mu\mu}(\omega_2,\omega_1)\big) G^{p_2}_c(\omega_2)G^{p_1}_c(\omega_1)\\& \ \ \ \ \ \ \ \ \ \ \ \ \ \ \ \ \ \ \  \ \ \ \ \ +\frac{g^2_0}{\sqrt{\omega_{p_1}\omega_{p_2}}}\int\frac{d\omega}{2\pi}\frac{\langle j_{d,\mu\mu}(\omega)\rangle e^{-i\omega t}}{\omega-(\omega_{p_1}+\omega_{p_2})+i(\gamma_{p_1}+\gamma_{p_2})}.
    \end{aligned}\label{Eq:F3}
\end{equation}
As can be seen, when we combine results we substitute Eq.\eqref{Eq:F2} and \eqref{Eq:F3} into \eqref{Eq:F1} the second terms in the inhomogeneous contributions cancel the corresponding terms in the homogeneous solution. As a result, we obtain for $\tau>0$
\begin{equation}
        \begin{aligned}
            \langle a^{\dagger}_{\mu,{p_1}}(t)a_{\mu,p_2}(t+\tau)\rangle&=\frac{g^2_0}{\sqrt{\omega_{p_1}\omega_{p_2}}}\int\frac{d\omega_1}{2\pi}\int\frac{d\omega_2}{2\pi}e^{-i(\omega_1+\omega_2)t}\langle j(\omega_1)j(\omega_2)\rangle G^{p_2}_c(\omega_2)G^{p^*_1}_c(-\omega_1)e^{-i\omega_2\tau}; \\
            \langle a_{\mu,{p_1}}(t)a_{\mu,p_2}(t+\tau)\rangle&=\frac{g^2_0}{\sqrt{\omega_{p_1}\omega_{p_2}}}\int\frac{d\omega_1}{2\pi}\int\frac{d\omega_2}{2\pi}e^{-i(\omega_1+\omega_2)t}\big(\langle j(\omega_1)j(\omega_2)\rangle+\omega_1\sigma_{\mu\mu}(\omega_2,\omega_1)\big) G^{p_2}_c(\omega_2)G^{p_1}_c(\omega_1)e^{-i\omega_2\tau}\\&-\frac{g^2_0}{\sqrt{\omega_{p_1}\omega_{p_2}}}\int\frac{d\omega_1}{2\pi}\int\frac{d\omega_2}{2\pi}e^{-i(\omega_1+\omega_2)t}\omega_1\sigma_{\mu\mu}(\omega_2,\omega_1)G^{p_2}_c(\omega_2)G^{p^*_1}_c(\omega_1)e^{-i\omega_2\tau}
        \end{aligned}, \label{Eq:F4}
\end{equation}
$\sigma_{\mu\nu}(\omega_1,\omega_2)=(i/\omega_2)(\chi_{\mu\nu}(\omega_1,\omega_2)+\langle j_{d,\mu\nu}(\omega_1+\omega_2)\rangle)$ is conductivity tensor. Here, we can observe an important difference between the  normal and anomalous correlations, namely the latter contains a diamagnetic term that acts as a parametric amplifier and can produce squeezing. In practice, we are not interested in the full form of Eq.\eqref{Eq:F4}, we instead consider the frequency-resolved correlations averaged over one laser period $\int d\tau e^{i\omega \tau}\overline{\langle A(t)B(t+\tau)\rangle}^t$ to determine the far-field properties. After some simple algebra we obtain
\begin{equation}
    \begin{aligned}
        \int\limits_{-\infty}^{\infty} d\tau e^{i\omega \tau}\overline{\langle a^{\dagger}_{\mu,{p_1}}(t)a_{\mu,p_2}(t+\tau)\rangle}^t&=\frac{g^2_0}{\sqrt{\omega_{p_1}\omega_{p_2}}}G^{p_2}_c(\omega)G^{p^*_1}_c(\omega)\left(|\langle j_{\mu}(\omega)\rangle|^2+D_{\mu\mu}(-\omega,\omega)\right); \\ 
        \int\limits_{-\infty}^{\infty} d\tau e^{i\omega \tau}\overline{\langle a_{\mu,{p_1}}(t)a_{\mu,p_2}(t+\tau)\rangle}^t&=\frac{g^2_0}{\sqrt{\omega_{p_1}\omega_{p_2}}}G^{p_2}_c(\omega)G^{p_1}_c(-\omega)\Big(|\langle j_{\mu}(\omega)\rangle|^2+D_{\mu\mu}(-\omega,\omega)-\omega\sigma_{\mu\mu}(\omega,-\omega)\Big)\\&
        +\frac{g^2_0}{\sqrt{\omega_{p_1}\omega_{p_2}}}G^{p_2}_c(\omega)G^{p^*_1}_c(-\omega)\omega\sigma_{\mu\mu}(\omega,-\omega)
    \end{aligned}\label{Eq:F5}
\end{equation}
where $D_{\mu,\nu}(\omega_1,\omega_2)=\int dt\int dt'e^{i\omega_1t}e^{i\omega_2t'}D_{\mu,\nu}(t,t')$. In the above equation,we have separated the coherent contribution, which scales as $N^2$, which states for coherent type of emission, and the incoherent part that grows as $N$. 

In Appendix~\ref{app:D2}, we obtain the Wigner representation for the correlation matter tensor, so it is natural to obtain the Fourier-transformed tensor from its Wigner representation as
\begin{equation}
\begin{aligned}
    D_{\mu,\mu}(-\omega_1,\omega_2)&=\int dt\int dt' e^{-i\omega_1t+i\omega_2t'}D_{\mu,\mu}(t,t')=\int dt e^{-i(\omega_1-\omega_2)t}D_{\mu,\mu}\left(t,-\frac{\omega_1+\omega_2}{2}\right).   
\end{aligned}\label{Eq:F6}
\end{equation}

Moreover, we need to consider frequency-resolved correlations. With the help of Eq.\eqref{Eq:F4} it is possible to immediately derive
\begin{equation}
    \begin{aligned}
        \langle a^{\dagger}_{\mu,p_1}(\omega_1)a_{\mu,p_2}(\omega_2)\rangle&=\frac{g^2_0}{\sqrt{\omega_{p_1}\omega_{p_2}}}G^{p^*_1}_c(\omega_1)G^{p_2}_c(\omega_2)\left(\langle j_{\mu}(-\omega_1)\rangle\langle j_{\mu}(\omega_2)\rangle+D_{\mu\mu}(-\omega_1,\omega_2)\right);\\
        \langle a_{\mu,p_1}(\omega_1)a_{\mu,p_2}(\omega_2)\rangle&=\frac{g^2_0}{\sqrt{\omega_{p_1}\omega_{p_2}}}G^{p_1}_c(\omega_1)G^{p_2}_c(\omega_2)\Big(\langle j_{\mu}(\omega_1)\rangle\langle j_{\mu}(\omega_2)\rangle+D_{\mu\mu}(\omega_1,\omega_2)+\omega_1\sigma_{\mu\mu}(\omega_2,\omega_1) \Big)\\&
        -\frac{g^2_0}{\sqrt{\omega_{p_1}\omega_{p_2}}}G^{p_2}_c(\omega_2)G^{p^*_1}_c(\omega_1)\omega_1\sigma_{\mu\mu}(\omega_2,\omega_1).
    \end{aligned}\label{Eq:F7}
\end{equation}  
\end{widetext}

\subsection{Far-field statistics}
\label{app:H}
The simplest statistic that characterizes the far fields is the intensity spectrum of the outgoing field, which can be calculated by studying the correlation function of the outgoing fields at two different times. 
\begin{equation}
\begin{aligned}
S(\omega,t)&=\int\limits_{-\infty}^{\infty}d\tau e^{i\omega \tau}\langle a^{\mu^{\dagger}}_{out}(t)a^{\mu}_{out}(t+\tau)\rangle.    
\end{aligned}\label{Eq:H1}
\end{equation}
Because the dephasing times are generally longer than one laser period, we average over one laser period and use the input-output relation in Eq.~\eqref{Eq:B1} to rewrite Eq.~\eqref{Eq:H1} as
\begin{equation}
\begin{aligned}
    &\overline{S}(\omega)\approx\langle a^{\mu^{\dagger}}_{out}(\omega)a^{\mu}_{out}(\omega)\rangle\\&=\left[|\langle j_{\mu}(\omega)\rangle|^2+D_{\mu,\mu}(-\omega,\omega)\right]\left|\sum\limits_{m}\sqrt{\frac{2\gamma_mg^2_0}{\omega_m}}G^m_{c}(\omega)\right|^2,
\end{aligned}\label{Eq:H2}
\end{equation}
where the sum over the mode index $m$ defines the multi-mode transfer function $F(\omega)\equiv\sum_mg_0G^m_c(\omega)/\sqrt{Q^m_c}$. In practice, this transfer function acts as a filter that suppresses nonresonant emission. The first coherent term can be obtained even in the semiclassical regime and it generally determines the dominant contribution to the HHG energy spectrum. However, as we know from the previous section, aall Ohmic contributions coming from the back-action cavity force are included in the first, coherent term, which brings its calculation beyond the standard semiclassical approach. Meanwhile, the last dispersive term adds an incoherent part to the total radiation.

Another photon statistic that  characterizes nontrivial correlations is the frequency-resolved second-order correlation function which can be determined as 
\begin{widetext}
\begin{equation}
\begin{aligned}
    g_2(\omega_1,\omega_2,\tau)_{\mu}&=\frac{\overline{\langle A^{\mu^{\dagger}}_{out}(\omega_1,t)A^{\mu^{\dagger}}_{out}(\omega_2,t+\tau)A^{\mu}_{out}(\omega_2,t+\tau)A^{\mu}_{out}(\omega_1,t)\rangle}^t}{\overline{\langle A^{\mu^{\dagger}}_{out}(\omega_1,t)A^{\mu}_{out}(\omega_1,t)\rangle}^{t}\overline{\langle A^{\mu^{\dagger}}_{out}(\omega_2,t+\tau)A^{\mu}_{out}(\omega_2,t+\tau)\rangle}^{t}}, \\ 
    A^{\mu}_{out}(\omega,t)=&\int \frac{d\omega'}{2\pi}h_{\omega}(\omega')a^{\mu}_{out}(\omega')e^{-i\omega't}
\end{aligned}\label{Eq:H4}
\end{equation}
\end{widetext}
where $h(\omega')$ is a narrow filter profile centered at $\omega$. In what follows, we consider only the simplest case, namely  $h(\omega')=2\pi\delta(\omega-\omega')$. Because of the particular choice of the filter function, the obtained $g_2(\omega)$ does not depend on $\tau$. Moreover, because of the low temperature there is no additional term due to the thermal bath and hence one can express all outgoing field operators $a_{out}$ in terms of their intracavity
counterparts and hence everything can be determined via intracavity field statistics. This second-order correlation function characterizes correlations between modes $\omega_1$ and $\omega_2$.  As a result, after some algebra we obtain
\begin{equation}
    \begin{aligned}
g_2(\omega_1,&\omega_2)_{\mu}=1+\frac{\left|1+\frac{D_{\mu\mu}(-\omega_1,\omega_2)}{\langle j_{\mu}(-\omega_1)\rangle\langle j_{\mu}(\omega_2)\rangle}\right|^2-1}
{(s(\omega_2)+1)(s(\omega_1)+1)}\\&+\frac{\left|1+\frac{D_{\mu\mu}(\omega_1,\omega_2)}{\langle j_{\mu}(\omega_1)\rangle\langle j_{\mu}(\omega_2)\rangle}+\frac{\omega_1\sigma_{\mu\mu}(\omega_2,\omega_1)\rangle}{\langle j_{\mu}(\omega_1)\rangle\langle j_{\mu}(\omega_2)\rangle}\right|^2-1}{(s(\omega_2)+1)(s(\omega_1)+1)}  ,\end{aligned}\label{Eq:H5}
\end{equation}    
where $s(\omega)=S_{incoh}(\omega)/S_{coh}(\omega)$. One can notice that the obtained correlation function is independent of the cavity parameters at leading order $g^2_0$. Hence, this result is universal and the corresponding free-space HHG result is recovered in this approximation. As can be seen, in the case of $\omega_1=\omega_2$, i.e. self-correlation, this quantity scales as $\sim1+\text{O}(1/N)$ and hence in general the photon statistics are close to Poissonian. Nevertheless, due to incoherent contributions we might expect corrections to the field statistics of the total radiation. The first contribution in Eq.\eqref{Eq:H5} is due to the normal correlation function. From its structure, it is clear that it causes the system to acquire super-Poissonian statistics when $s(\omega)>0$, because the value is greater than 1. The second contribution is generated by the anomalous correlation. Depending on its magnitude and phase, it can push the system to super- or sub-Poissonian statistics.

In spite of the fact that so far we have not demonstrated any quantum features of radiation, the degree of squeezing can reveal the quantum properties of the field. As stated in the main text, one can characterize the degree of squeezing in Eq.~\eqref{Eq:degree_sq} by the variance of the rotated quadrature, which can be introduced as 
\begin{equation}
    \hat{X}_{\mu}(t,\theta)=\frac{1}{2}\left(a^{\mu}_{out}(t)e^{-i\theta}+a^{\mu^{\dagger}}_{out}(t)e^{i\theta}\right),\label{Eq:H6}
\end{equation}
where $\theta$ is the quadrature angle. As a result, one can measure the squeezing spectrum via homodyne spectroscopy as 
\begin{equation}
\Delta\hat{X}_{\mu,\omega}(\theta) = \frac{1}{4}+\int\limits_{-\infty}^{\infty}d\tau e^{i\omega\tau}\overline{\langle :\hat{X}_{\theta}(t),\hat{X}_{\theta}(t+\tau):\rangle}^t,\label{Eq:H7}
\end{equation}
where $\langle:..:\rangle$ is a normal operator ordering operation. After some calculations, we can rewrite it as 
\begin{equation}
\begin{aligned}
    \Delta\hat{X}_{\mu,\omega}(\theta)=\frac{1}{4}&+\frac{1}{2}\left[S_{sym}(\omega)+\text{Re}\left(M(\omega)e^{-2i\theta}\right)\right],
\end{aligned}\label{Eq:H8}
\end{equation}
where $S_{sym}(\omega)=(\overline{S}_{incoh}(\omega)+\overline{S}_{incoh}(-\omega))/2$ is the symmetrized power spectral density and $M(\omega)$ is the anomalous spectral density, which is given by 
\begin{equation}
\begin{aligned}
    M(\omega)=\big(D_{\mu\mu}(-\omega,\omega)-\omega\sigma_{\mu\mu}(\omega,-\omega)\rangle\big)F(\omega)F(-\omega).    
\end{aligned}\label{Eq:H9}
\end{equation}
As can be noticed, because of the properties of coherent states, this expression does not depend on coherent fields at all. Its value is determined by the noise correlation functions. This in turn proves our statement that noise correlation functions that are determined by current-current fluctuations in a solid are responsible for quantum light generation. Taking into account the relation between the spectral noise function and the generalized noise correlation function and optimizing over all angles, one can obtain the following
\begin{equation}
\begin{aligned}
    \Delta\hat{X}_{\mu,\omega}(\theta_{opt})&=\frac{1}{4}+\frac{1}{2}\left(S_{sym}(\omega)-|M(\omega)|\right).
\end{aligned}\label{Eq:H10}
\end{equation}
Therefore, the outgoing radiation is squeezed if $\Delta\hat{X}_{\mu,\omega}(\theta_{opt})<1/4$ or $S_{sym}(\omega)<|M(\omega)|$, which means that the symmetrized power spectral density must be smaller than the magnitude of the anomalous spectral density in order to obtain squeezed emitted light.

\bibliography{HHG.bib}
\end{document}